\documentclass[fleqn,usenatbib]{mnras}

\usepackage[T1]{fontenc}
\usepackage{ae,aecompl}

\usepackage{graphicx}	
\usepackage{amsmath}	
\usepackage{amssymb}	
\usepackage{gensymb}
\usepackage{mathrsfs}
\usepackage{txfonts, mathptmx}
\usepackage{dblfloatfix}
\usepackage{bm}
\usepackage{footmisc}
\usepackage{relsize}
\usepackage{booktabs}
\usepackage{hyperref}
\usepackage{units}
\usepackage{subfiles}
\usepackage{rotating}
\usepackage{booktabs}
\usepackage{caption}
\usepackage{longtable}
\usepackage{pdflscape} 

\usepackage{xspace}
\usepackage{color,soul}
\usepackage[dvipsnames,svgnames]{xcolor}

\newcommand{\kms}{\, \mathrm{km} \, \mathrm{s}^{-1}}
\newcommand{\junit}{\, {\rm GeV^{2} \, cm^{-5} } }
\newcommand{\dunit}{\, {\rm GeV \, cm^{-2} } }
\newcommand{\kpc}{\, \mathrm{kpc}}

\newcommand{\slos}{\sigma_{\mathrm{los}}}

\title[J and D-Factor Scaling]{Scaling Relations for Dark Matter Annihilation and Decay Profiles in Dwarf Spheroidal Galaxies}
 
\author[A. B. Pace \& L. E. Strigari]{
Andrew B. Pace,$^{1, 2}$ \thanks{E-mail: apace@tamu.edu }
Louis E. Strigari$^{1}$\\
$^1$George P. and Cynthia Woods Mitchell Institute for Fundamental Physics and Astronomy, and \\
Department of Physics and Astronomy, Texas A\&M University, College Station, TX 77843, USA\\
$^2$ Mitchell Astronomy Fellow \\ 
}
 
\date{Accepted XXX. Received YYY; in original form ZZZ}

\pubyear{2018}

\begin{document}
\label{firstpage}
\pagerange{\pageref{firstpage}--\pageref{lastpage}}
\maketitle

\begin{abstract}
Measuring the dark matter distribution in dwarf spheroidal galaxies (dSphs) from stellar kinematics is crucial for indirect dark matter searches, as these distributions set the fluxes for both dark matter annihilation (J-Factor) and decay (D-Factor). Here we produce a compilation of J and D-Factors for dSphs, including new calculations for several newly-discovered Milky Way (MW) satellites, for dSphs outside of the MW virial radius, and for M31 satellites. From this compilation we test for scaling relations between the J and D-factors and physical properties of the dSphs such as the velocity dispersion ($\slos$), the distance ($d$), and the stellar half-light radius ($r_{1/2}$). We find that the following scaling relation minimizes the residuals as compared to different functional dependencies on the observed dSphs properties
$J(0.5\degree) =  10^{17.87} \left(\slos/5\,\kms\right)^4 \left(d / 100\,\kpc\right)^{-2}\left(  r_{1/2}/100 \,{\rm pc} \right)^{-1}$. We find this relation has considerably smaller scatter as compared to the simpler relations that scale only as $1/d^2$. We further explore scalings with luminosity ($L_V$), and find that the data do not strongly prefer a scaling including $L_V$ as compared to a pure $1/d^2$ scaling. The scaling relations we derive can be used to estimate the J-Factor without the full dynamical analysis, and will be useful for estimating limits on particle dark matter properties from new systems that do not have high-quality stellar kinematics. 
\end{abstract}


\section{Introduction}
Dwarf spheroidal galaxies (dSphs) have proven to be ideal targets for gamma-ray searches from particle dark matter annihilation or decay~\citep{Conrad2015JETP..121.1104C}. The Fermi-LAT combined limits from the dSphs with well-measured stellar kinematics place strong bounds on the dark matter annihilation cross section, with sensitivity to cosmologically-motivated models in the mass range $\sim 10-100$ GeV \citep{FermiLATCollaboration2010ApJ...712..147A, Geringer-Sameth2011PhRvL.107x1303G, FermiLATCollaboration2011PhRvL.107x1302A, FermiLATCollaboration2014PhRvD..89d2001A, Geringer-Sameth2015PhRvD..91h3535G, FermiLATCollaboration2015PhRvL.115w1301A}. More massive dark matter particles, $\sim 100$ GeV - $100$ TeV, are now being probed by ground-based gamma-ray observatories \citep{HESS2014PhRvD..90k2012A, MAGIC2016JCAP...02..039M, HAWCCollaboration2017arXiv170601277A, VERITAS2017PhRvD..95h2001A}.

The gamma-ray flux from dark matter annihilation or decay depends on the distribution of dark matter within the system (the astrophysical component) and the annihilation cross section or decay rate which determines how the dark matter particles convert into standard model particles (the particle physics component). The astrophysical component that governs the annihilation and decay fluxes are commonly referred to as the J and D-Factors, which are, respectively, line-of-sight integrals over the dark matter distribution squared and the dark matter distribution within a dSph. 

The measured stellar kinematics of dSphs determine the dark matter density distribution, and thereby the J and D-factors. There is a large volume of literature on computing the J and D-Factors. The standard methodology for computing these factors for the dSphs involves combining the spherical Jeans equation with a Bayesian likelihood analysis to constrain model parameters~\citep{Evans2004PhRvD..69l3501E, Strigari2007PhRvD..75h3526S, Strigari2008ApJ...678..614S, Martinez2009JCAP...06..014M, Charbonnier2011MNRAS.418.1526C, Bonnivard2015MNRAS.446.3002B, Geringer-Sameth2015ApJ...801...74G, Bonnivard2015MNRAS.453..849B}. 
The dynamical analyses indicate that the J and D-factors are in particular well-constrained for angular scales subtended by the approximate half-light radius of the stellar system~\citep{Walker2011ApJ...733L..46W}, which corresponds to an angular scale $\sim 0.1-0.5$ degrees for a typical dSph. The spherical Jeans formalism may be extended to consider an axisymmetric stellar distribution~\citep{Hayashi2016MNRAS.461.2914H, Klop2017PhRvD..95l3012K}. The results from the spherical Jeans modeling are consistent with simple analytic  relations \citep{Evans2016PhRvD..93j3512E} including axisymmety~\citep{Sanders2016PhRvD..94f3521S}. The astrophysical properties of the global population of dSphs may be used to better constrain the J-factors for individual systems, for example using Bayesian hierarchical modeling \citep{Martinez2015MNRAS.451.2524M}. The Bayesian techniques used in the aforementioned analyses can be compared to a Frequentist-based method, which is more compatible with the statistical modeling used in the Fermi-LAT gamma-ray analysis \citep{Chiappo2017MNRAS.466..669C}.

\par In recent years, many new dSphs have been discovered in the Sloan Digital Sky Survey (SDSS), PanSTARRS \citep{Laevens2015ApJ...802L..18L, Laevens2015ApJ...813...44L}, and the Dark Energy Survey (DES) \citep{Bechtol2015ApJ...807...50B, Koposov2015ApJ...805..130K, Drlica-Wagner2015ApJ...813..109D}. 
These systems are sufficiently faint that it is difficult to obtain spectroscopy on a large sample of their member stars. Because of these small data samples, and the susceptibility to systematic uncertainties, the velocity dispersion, and therefore the mass and J-factor, may easily be systematically overestimated if the velocity dispersion is based on only a handful of stars~\citep{Bonnivard2016MNRAS.462..223B}. However, given the importance of the dSphs to dark matter searches, going forward it will be increasingly important to obtain reliable estimates for the J and D-factors, especially for systems without kinematic measurements.

\par Because many newly-discovered dSphs lack measured stellar kinematics, previous studies have appealed to scaling relations between the J-factor and observable quantities to extract dark matter limits from them. For example \citet{Drlica-Wagner2015ApJ...809L...4D, Albert2017ApJ...834..110A} consider a simple model in which the J-factor scales as the inverse of the distance squared to the dSph. This scaling was motivated by the fact that the dSphs have similar integrated dark matter masses, despite having a wide range of luminosities \citep{Strigari2008Natur.454.1096S, Walker2009ApJ...704.1274W, Wolf2010MNRAS.406.1220W}.  As even more dSphs are discovered, it will be increasingly important to identify the most optimal scaling relations between the J and D-factors and the observable properties of the dSphs. 

\par In this paper, we compute the J and D-factors from the dSphs with measured stellar kinematics, and perform new calculations for several  systems that do not have published measurements of these quantities. We include not only the Milky Way (MW) satellites, but also dSphs that are satellites of M31, and are in the local field (LF), not bound to either the MW or M31. We perform a statistical analysis to determine the observable quantities, in particular the line-of-sight velocity dispersion ($\slos$), distance ($d$), and stellar scale ($r_{1/2}$), that the J and D-factors scale with. We determine the appropriate scaling relations, and quantify the residuals obtained from these relations. Our statistical analysis builds on the analytic work of \citet{Evans2016PhRvD..93j3512E}, who work out relations for the J-factors in terms of parameters that describe the dark matter halo, such as the scale density and the scale radius. 

We structure of this paper as follows. In Section~\ref{section:data}, we summarize the dSph properties and data sources.  In Section~\ref{section:method}, we present our dynamical framework for determining the dark matter distributions and subsequent J and D-Factor calculations.  In Section~\ref{section:results}, we present results for the galaxy sample, present scaling relations for the J and D-Factors, and discuss some systematics. In Section~\ref{section:conclusion}, we conclude.

\section{Data} 
\label{section:data}
To determine the J and D-factors, we compile a large sample of dSph spectroscopic data from the literature. The sample properties are summarized in Table~\ref{data_table} and includes the distance ($d$), azimuthally averaged half-light radius ($r_{1/2}$), line-of-sight velocity dispersion ($\slos$), absolute magnitude (${\rm M_V}$), and number of spectroscopic members (N).  
We have opted not to combine spectroscopic samples due to potential zero-point offsets in heliocentric velocity between different telescopes/instruments or mis-estimation of velocity errors. We use the Gaussian likelihood given in \citet{Walker2006AJ....131.2114W} to compute the average velocity, $\overline{V}$,  and $\slos$.  For $\slos$ we assume a Jefferys prior with range $-2 < \log_{10}{\slos} < 2$. The resulting values of $\slos$  are listed in Table~\ref{data_table}.

For galaxies not mentioned in the reminder of the section the membership was determined in the original spectroscopic study (citations are listed in number of stars column in Table~\ref{data_table}). 
The data for Carina, Fornax, Sculptor, and Sextans is from  \citet{Walker2009AJ....137.3100W}, the Draco data is from ~\cite{Walker2015MNRAS.448.2717W}, and the Ursa Minor data is from M. Spencer et al. in prep (Matt Walker private communication).
For all 6 galaxies, we select member stars by applying a cut at $p_i > 0.95$, where $p_i$ is the membership probability  determined from the expectation maximization method \citep{Walker2009AJ....137.3109W}.   
For Leo I \citep{Mateo2008ApJ...675..201M} and  Leo II \citep{Spencer2017ApJ...836..202S} we use a $3-\sigma$ clipping algorithm to select members.

For Bo\"{o}tes I, we use the 37 member ``Best'' sample in \citet{Koposov2011ApJ...736..146K}.  We additionally explored a larger subset of this sample and found consistent results.
We have made slight membership changes to the~\citet{Simon2007ApJ...670..313S} data for the following dSphs: Canes Venatici I, Coma Berenices, Leo IV, Ursa Major I, and Ursa Major II.  
We have removed RR Lyrae from the following (identified after the original spectroscopic analysis): Canes Venatici I \citep[5 RR Lyrae identified in][]{Kuehn2008ApJ...674L..81K}, Coma Berenices \citep[1 star;][]{Musella2009ApJ...695L..83M}, Leo IV \citep[1 star;][]{Moretti2009ApJ...699L.125M}, and Ursa Major I \citep[3 stars;][]{Garofalo2013ApJ...767...62G}. 
RR Lyrae are pulsating stars with variable velocity; these stars are dSph members, however, without additional phase information we cannot determine the star's bulk velocity.
In Ursa Major II, a foreground dwarf has been removed that was identified in follow-up high-resolution spectroscopy analysis \citep{Frebel2010ApJ...708..560F}.
These removals do not have a large impact as the removed stars tend to have large error bars. 
For Segue 1, we used the Bayesian membership probability ($p_i>0.8$) to identify Segue 1 members \citep{Simon2011ApJ...733...46S}.

The Hercules data set has 18 members \citep{Aden2009A&A...506.1147A} but we have removed one member that was later identified as a spectroscopic binary\footnote{The binary star's center of mass velocity was identified but there is a zero point offset between the different studies.} \citep{Koch2014ApJ...780...91K}.  
Regarding Hercules it is also worth noting that several studies indicate that this galaxy may be undergoing tidal disruption~\citep[e.g.][]{Deason2012MNRAS.425L.101D, Kupper2017ApJ...834..112K, Garling2018ApJ...852...44G}.
For Leo V, we use the 8 star data set from \cite{Collins2017MNRAS.467..573C}.  We note that this analysis argued that Leo V contains a large velocity gradient,  a potential indication of tidal disruption. 
For Willman 1, we use the 40 likely members selection but note that the dynamical state of this system is unclear \citep{Willman2011AJ....142..128W}. Furthermore, Willman 1 member selection is difficult due to the overlap in heliocentric velocity with the MW field stars \citep{Siegel2008AJ....135.2084S, Willman2011AJ....142..128W}.  
The satellites Segue 2 \citep{Kirby2013ApJ...770...16K} and Triangulum II \citep{Kirby2017ApJ...838...83K} have also been argued to have undergone tidal disruption due to their offset on the stellar mass-luminosity relation \citep{Kirby2013ApJ...779..102K}.
We also note that the satellite, Tucana III, contains tidal tails \citep{Drlica-Wagner2015ApJ...813..109D, Shipp2018arXiv180103097S}.

The data for  Grus I and Tucana II is from~\citet{Walker2016ApJ...819...53W}.  
We find that simply weighting the stars by their reported mean membership values gives a $\slos$ result that disagrees with the values reported in~\citet{Walker2016ApJ...819...53W}. 
We speculate that this disagreement is due to our lack of a Milky Way background model; there are several stars with large errors on the dSph membership probabilities that could be driving the dispersions apart. In order to reproduce the \citet{Walker2016ApJ...819...53W} values, we exclude (include) the two stars in Grus I (Tucana II) with large membership errors but non-zero membership giving a sample of 5 (10) stars.

We include a handful of more distant objects in our compilation including five M31 satellites and four local field (LF) objects.  
We selected the 5 Andromeda satellites with the largest sample size from the SPLASH survey \citep{Tollerud2012ApJ...752...45T}. 
Following the SPLASH team, dSph stars are selected with a membership probability cut of $p_i>0.1$.
The LF objects we include here are: And XVIII \citep[][$d_{\rm M31} - d_{\rm And \, XVIII} \approx 600 \kpc$]{Tollerud2012ApJ...752...45T}, Cetus \citep{Kirby2014MNRAS.439.1015K}, Eridanus II \citep{Li2017ApJ...838....8L}, and Leo T \citep{Simon2007ApJ...670..313S}.  
We did not consider any other local group dSph/dIrr as our dynamical modeling assumptions are not appropriate;  they  contain either low mass-to-light ratios \citep[e.g. Aquarius, Leo A, NGC 6822,][]{Kirby2014MNRAS.439.1015K},  large gas components,  disk-like stellar components,   peculiar kinematics \citep[e.g. Tucana,][]{Fraternali2009A&A...499..121F}, and/or  stellar rotation \citep[e.g. Pegasus and Phoenix,][]{Kirby2014MNRAS.439.1015K, Kacharov2017MNRAS.466.2006K}.

\section{Methods}
\label{section:method}

\par In this section we briefly outline the method for calculating the J and D-factors. To facilitate comparison with previous results, we follow standard treatments in the literature~\citep[e.g.][]{Evans2004PhRvD..69l3501E, Strigari2007PhRvD..75h3526S, Strigari2008ApJ...678..614S, Martinez2009JCAP...06..014M, Charbonnier2011MNRAS.418.1526C, Bonnivard2015MNRAS.446.3002B, Geringer-Sameth2015ApJ...801...74G, Bonnivard2015MNRAS.453..849B}, and refer to these papers for more details and discussion of systematics.

\par The J-Factor (annihilation factor) is an integral over the line-of-sight and solid angle of the dark matter density squared: 
\begin{equation}
J(\theta_{\mathrm{max}}) = \underset{\mathrm{los}}{\iint} \rho_{\mathrm{DM}}^2 (r) \, \mathrm{d}\ell \mathrm{d}\Omega\,,
\end{equation}
\noindent where $\rho_{\mathrm{DM}}$ is the dark matter density profile, $\ell$ is the  line-of-sight direction and $\Omega$ is the solid angle of radius $\theta_{\mathrm{max}}$.  The relationship between $r$ and $\ell$ is: $r^2 = \ell^2 + d^2 - 2 \ell d \cos{\theta}$ and ${\rm d}\Omega=2\pi \sin{\theta}{\rm d}\theta$.
The limits of the $\ell$ integration are: $\ell_{\pm}= d \cos{\theta} \pm \sqrt{r_t^2 - d^2 \sin^2{\theta}}$ and $r_t$ is the dark matter tidal radius.
The D-Factor (decay factor) is:
\begin{equation}
D(\theta_{\mathrm{max}}) = \underset{\mathrm{los}}{\iint} \rho_{\mathrm{DM}} (r) \, \mathrm{d}l \mathrm{d}\Omega\,.
\end{equation}

\par For both the J and D-factor, the key quantity to determine from the stellar photometry and kinematic data is  $\rho_{\mathrm{DM}}$. The dark matter density is constrained through the spherical Jeans equation, 
\begin{equation}
\frac{\mathrm{d}\nu \sigma_r^2}{\mathrm{d}r} + \frac{2}{r} \beta(r) \nu \sigma_r^2 +  \frac{\nu \mathrm{G} M(r)}{r^2} = 0\,,
\label{eq:jeans}
\end{equation}
where $\nu$ is the tracer (stellar) density, $\sigma_r^2$ is the radial velocity dispersion, $\beta(r) = 1 - \frac{\sigma_t^2}{2 \sigma_r^2}$ is the velocity anisotropy, and $M(r)$ is the mass profile.  Since the dSphs have large mass-to-light ratios (with the only exception being the central region of Fornax), we assume that the mass profile is entirely dominated by the dark matter halo.

\par To compare $\sigma_r^2$ to observed line-of-sight velocity data it needs to be projected into the line-of-sight direction.  The projection is:
\begin{equation}
\Sigma(R) \sigma_{\mathrm{los}}^2 (R) = 2 \int_R^{\infty}\mathrm{d}r \, \left[1 - \beta(r) \frac{R^2}{r^2} \right] \frac{r \, \nu \sigma_r^2(r)}{\sqrt{r^2 - R^2}},
\end{equation}
\noindent where $\sigma_{\mathrm{los}}$ is the line-of-sight velocity dispersion and $\Sigma$ the projected tracer density.  We model the stellar profile with a Plummer (\citeyear{Plummer1911MNRAS..71..460P}) model.  The 3D distribution is: 

\begin{equation}
\nu_{Plummer} (r) = \frac{3}{4 \pi r_p^3} \frac{1}{\left(1 + (r/r_p)^2 \right)^{5/2}}. 
\end{equation}

\noindent and the projected (2D) distribution is:

\begin{equation}
\Sigma_{Plummer} (r) = \frac{1}{ \pi r_p^2} \frac{1}{\left(1 + (R/r_p)^2 \right)^{2}}. 
\end{equation}

\noindent The Plummer profile has a scale radius, $r_p$, which is equivalent to the stellar half-light radius.
The dSphs have varying ellipticity \citep[$\epsilon\approx0.1-0.7$][]{McConnachie2012AJ....144....4M}; to approximate spherical symmetry,  we use azimuthally\footnote{Sometimes referred to as the geometric half-light radius.} averaged half-light radii: $r_{1/2} = r_{\rm half, \, azimuthal} = r_{\rm half, \,major} \sqrt{1-\epsilon}$.  When referring to the half-light radius ($r_{1/2}$), we are referring to the azimuthally averaged value. Note that it is possible to consider more generalized models than the Plummer profile \citep{Strigari2010MNRAS.408.2364S}, however for the majority of our sample (ultra-faint dwarf galaxies) Plummer profiles are adequate to describe them.
We parameterize the dark matter distribution as a NFW profile \citep{Navarro1996ApJ...462..563N}:

\begin{equation}
\rho_{DM} = \frac{\rho_s}{x(1+x)^2}\,,
\end{equation}

\noindent where $x=r/rs$, and the scale radii and density are $r_s$ and $\rho_s$ respectively.  We take the velocity anisotropy to be constant with radius, $\beta(r) = \beta_0$. Instead of a  prior in linear $\beta$ space, we parameterize the anisotropy in a symmetrized space; $\tilde{\beta} = \beta/\left(2-\beta\right)$ \citep[see Eq. 8 in][]{Read2006MNRAS.367..387R, Read2017MNRAS.471.4541R}.  The symmetrized parameterization uniformly favors radial and tangential orbits whereas the linear parameterization  preferentially samples tangential orbits\footnote{Another alternative anisotropy parameterization for equally favoring radial and tangential orbits is: $\beta^{\prime}=\log_{10}{\left(1-\beta \right)}$ \citep{Charbonnier2011MNRAS.418.1526C}. }.

To extract the model parameters from the data, we use an unbinned likelihood function  \citep{Strigari2008ApJ...678..614S, Martinez2009JCAP...06..014M, Geringer-Sameth2015ApJ...801...74G}:

\begin{equation}
\mathcal{L}_v (\mathcal{D}) = \prod_{i=1}^{N} \frac{p_i}{\sqrt{2 \pi (\sigma_{\mathrm{los}}^2 (R_i)+\sigma_{\epsilon, i}^2)}} \exp{\left[-\frac{1}{2} \frac{(V_i - \overline{V})^2}{\sigma_{\mathrm{los}}^2 (R_i) +\sigma_{\epsilon, i}^2}\right]},
\end{equation}

\noindent Here each data point is the radial position, line-of-sight velocity, velocity error, and membership probability; $\mathcal{D}_i= (R_i, V_i, \sigma_{\epsilon, i}, p_i)$.  For many data sets, there is  only member/non-member selection; the members are considered $p_i=1$ and the non-member discarded (see Section~\ref{section:data} and citations in the $N$ column of Table~\ref{data_table} for additional information).
$\overline{V}$ is the average heliocentric velocity of the dSph.

For the classical MW satellites with measured proper motions, we also account for the perspective motion affect outlined in \citet{Kaplinghat2008ApJ...682L..93K} and  \citet{Walker2008ApJ...688L..75W}.  
Briefly, for a large extended object (i.e. a ``classical'' dSph), only at the center of the object do the line-of-sight and z-directions exactly align.  A small component of the line-of-sight velocity comes from the net proper motion of the galaxy; this effect increases the further from the center.  Note this is generally a small effect, corresponding to $1-2\kms$.  To correct for this effect we follow the method outlined in the Appendix of \citet{Walker2008ApJ...688L..75W}. The proper motions (in ${\rm mas \, century^{-1} }$) of the classical satellites are as follows ($\mu_{\alpha} \cos{\delta}, \mu_{\delta}$)= ($22\pm9$, $15\pm9$) \citep[Carina;][]{Piatek2003AJ....126.2346P}, ($5.62\pm0.99$, $-16.75\pm1.0$) and ($2.96\pm2.09$, $-13.58\pm2.14$) \citep[Draco and Sculptor;][]{Sohn2017ApJ...849...93S}, ($47.6\pm4.5$, $-36.0\pm4.1$) \citep[Fornax;][]{Piatek2007AJ....133..818P}, ($-11.4\pm2.95$, $-12.56\pm2.93$) \citep[Leo I;][]{Sohn2013ApJ...768..139S}, ($-6.9\pm3.7$, $-8.7\pm3.9$) \citep[Leo II;][]{Piatek2016AJ....152..166P}, ($-40.9\pm5.0$, $-4.7\pm5.8$) \citep[Sextans;][]{Casetti-Dinescu2017arXiv171002462C} , and ($-50\pm17$, $22\pm16$) \citep[Ursa Minor;][]{Piatek2005AJ....130...95P}.  When available, we choose proper motion results based on data from the {\it Hubble Space Telescope} over ground based studies.  We do not include this effect for the ultra-faint satellites as they do not have measured proper motions and are less extended systems.

We determine the posterior distributions with the MultiNest sampling routine \citep{Feroz2008MNRAS.384..449F, Feroz2009MNRAS.398.1601F}.
We assume Gaussian priors based on literature measurements for the distance, half-light radius\footnote{In some literature fits an exponential model was used instead of the Plummer model assumed here.  To convert between Plummer and exponential scale radii we use: $r_{p} = 1.68\times r_{exp}$.}, ellipticity, and if applicable the proper motions. 
The literature measurements are summarized in Table~\ref{data_table} except for the proper motions which where summarized in the preceding paragraph.
 
To approximate Gaussianity, some parameter errors represent the average of the upper and lower error bars\footnote{We note that it is better to redraw the measured parameter distributions directly from the posterior distributions.  In most cases these are not available but they are from M31 distances \citep{Conn2012ApJ...758...11C} and M31 structural parameters \citep{Martin2016ApJ...833..167M}.  For additional discussion on this topic see  \citet{Martin2016ApJ...833..167M}.}.  
We assume a uniform prior range for $\overline{V}$: $-10 \leq \overline{V} - V_{\rm literature} \leq +10$\footnote{This prior range is expanded for several satellites with small sample sizes.}.
Jeffreys priors are assumed for the dark matter halo parameters:  $-2 \leq \log_{10}{r_s} \leq 1$, and $4 \leq \log_{10}{\rho_s} \leq 14$. 
We additionally impose the prior $r_s > r_{1/2}$ \citep[see Section 4.1 of ][and discussion in Section~\ref{section:halo_prior}]{Bonnivard2015MNRAS.446.3002B}.
The prior range for the anisotropy parameter is: $-0.95 \leq \tilde{\beta} \leq 1$.  
In summary we have 4 free parameters for the Jeans modeling ($\overline{V}$, $r_s$, $\rho_s$, and $\tilde{\beta}$) and 2-5 parameters with Gaussian priors to average over observational uncertainties ($d$, $r_p$, $\epsilon$, $\mu_{\alpha}$, $\mu_{\delta}$).

The dark matter tidal radius, $r_t$, is required for the J and D-Factor calculation.
We compute $r_t$ via: $r_t = \left[M_{sub}(r_t)/(2 - \mathrm{d}\ln{(M_{host}})/\mathrm{d}\ln{r}) M_{host}(d) \right]^{1/3} d$ \citep[Eq. 12 of ][]{Springel2008MNRAS.391.1685S}.
The MW and M31 host mass profiles are from \citet{Eadie2016ApJ...829..108E} and \citet{Sofue2015PASJ...67...75S} respectively.
The LF objects  systems have $r_t$ fixed to $r_t =25\kpc$.  
The LF $r_t$ is set based on the Jacobi radii of the classical MW satellites which have median $r_t$ values in the range $6.3-38.9 \kpc$.  Choosing a different value ($10~\kpc$ or $5~\kpc$) does not affect the J-Factor as the majority of the `signal/flux’ comes from the inner region of the satellite (this is not true for the D-Factor).  
Systems with an unresolved $\slos$ have $r_t$ fixed to $r_t=1\kpc$. 
We base the choice on the $r_t$ values of nearby ($d<60~\kpc$) ultra-faints.  The median $r_t$ of these systems range from $\sim2-6~\kpc$.  Only a small portion of the $r_t$ posterior of closest satellites falls below 1 kpc.

\section{Results and Discussion}
\label{section:results}

\subsection{Unresolved Velocity Dispersions}
\label{section:sigma}

\begin{figure}
\includegraphics[width=\columnwidth]{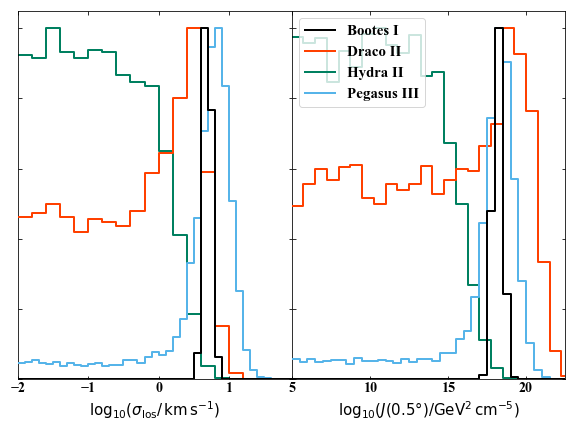}
\caption{Left: Example posteriors of four galaxies with different resolution of the velocity dispersion ($\slos$).  The resolved $\slos$ system is in black (Bo\"{o}tes~I), the green lines are a upper limit $\slos$ system (Hydra~II), and the orange (Draco~II) and light blue (Pegasus~III) contain large and small tails to $0-\slos$ respectively.  Right: The corresponding J-Factors at 0.5\degree.  All histograms are normalized to their maximum value to compare the shapes of the posterior.  
}
\label{fig:unresolved}
\end{figure}

There are several objects for which a non-negligible portion of the $\slos$  posterior tends to zero. 
For some systems the $\slos$ posteriors are entirely  upper limits \citep[Hydra~II, Segue~2, Triangulum~II, Tucana~III;][]{Kirby2015ApJ...810...56K, Kirby2013ApJ...770...16K, Kirby2017ApJ...838...83K, Simon2017ApJ...838...11S}, whereas others have large tails to zero-$\slos$ \citep[Draco~II, Leo~IV, Grus~I;][]{Martin2016MNRAS.458L..59M, Simon2007ApJ...670..313S, Walker2016ApJ...819...53W}, or small tails \citep[Leo V\footnote{In Leo V we find a similar small zero-$\slos$ tail in the 5 star data set from \citet{Walker2009ApJ...694L.144W}.}, Pegasus III, Pisces II;][]{Walker2009ApJ...694L.144W, Collins2017MNRAS.467..573C, Kim2016ApJ...833...16K, Kirby2015ApJ...810...56K}.  The tails are easily observed in $\log_{10}{\slos}$ space and may have been previously overlooked if only examined in linear space.
We have separated our sample into 4 groups based on the shape of the $\slos$ posterior; 
3 groups based on how much of the posterior tends to zero (unresolved, large zero-dispersion tail, and small zero-dispersion tail and 1 group with secure $\slos$ measurements.
The distinction between large and small tail is based on the ratio of the posterior of the tail to peak.  The small tail galaxies have a ratios $\approx5\%$ whereas the large tail objects have ratios in the range $\approx 40-60\%$.

For the non-resolved systems, we have expanded the prior range of the dark matter scale density ($0 < \log_{10}{\left(\rho_s/ {\rm M}_{\odot}\, {\rm kpc}^3 \right)} < 13$) to better illuminate the zero J-factor tail in the posterior (with the original prior only a small portion of the posterior corresponds to zero-$\slos$). Note that $\rho_s$ values in the expanded prior range are highly disfavored from cosmological N-Body simulations \citep{Springel2008MNRAS.391.1685S}.
Figure~\ref{fig:unresolved} shows $\slos$ (left) and J-Factor (right) posteriors for  representative galaxies from the four cases.  A non-resolution of all or part of $\slos$ implies a similar non-resolution in the J-Factor posterior and both have similar shapes.
Note that both posterior distributions will extend to even smaller values if the prior range is expanded. 
The lower limit of $\rho_s$ will set the inferred confidence intervals of the J and D-Factors and confidence intervals for objects without a resolved $\slos$ should be treated with caution.
For the subset of galaxies with zero-$\slos$ tails we quote maximum posterior values and error bars encompass the non-tail portion of the posterior in Table~\ref{data_table}.
For unresolved systems we quote an upper limit at the 95.5\% percentile confidence interval after applying a cut at $\log_{10}{\rho_s }>4$.  This allows for better comparison to the remainder of our sample and to previous studies. 
For the remainder of this work we only include systems with securely measured $\slos$. 

\subsection{Halo  Priors}
\label{section:halo_prior}

\begin{figure}
\includegraphics[width=\columnwidth]{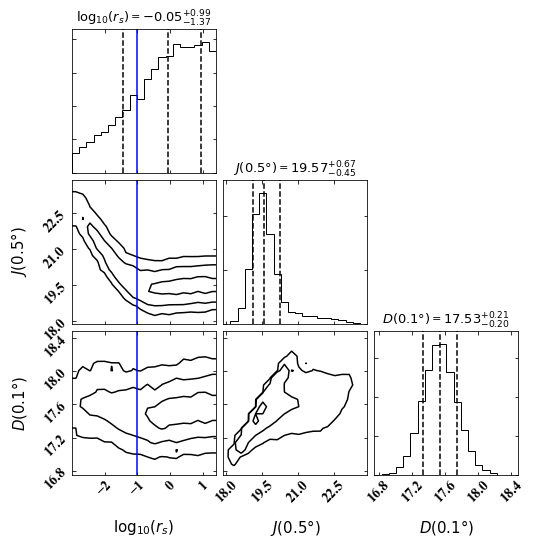}
\caption{Example of posterior without $r_s > r_{1/2}$ prior.  The corner plot compares $\log_{10}{r_s}$, $\log_{10}{J(0.5\degree)}$ and $\log_{10}{D(0.1\degree)}$ of Ursa Major II.  At small $r_s$, the J-Factor is systematically larger which is removed with our  $r_s > r_{1/2}$ prior. In the $r_s$ panels we display $r_{1/2}$ with a blue line.}
\label{fig:rsprior2}
\end{figure}

In our analysis we have imposed a prior $r_{s} > r_{1/2}$ \citep[as suggested by][]{Bonnivard2015MNRAS.446.3002B}.  To quantify the effect of this prior, we re-analyze several galaxies (Bo\"{o}tes~I, Carina~II, Reticulum~II, Ursa~Major~II, Draco, and Leo~II) without this prior and expand the $r_s$ prior range ($-3 < \log_{10}{\left( r_s/{\rm kpc} \right)} < 1.4$).  
While there is no change for the Draco values (the posterior favors larger $r_s$ values) all other galaxies show an increase in the J-Factor and most show a decrease in the D-Factor.  
This affect is generally larger in the ultra-faint galaxies as the halo properties are more likely to be prior dominated.
As an example, Figure~\ref{fig:rsprior2} shows a corner plot comparing the posteriors of $r_s$, $J(0.5\degree)$, and $D(0.1\degree)$ for Ursa~Major~II.
For  $r_s>r_{1/2}$, there is little to no $r_s$ trend with the J-Factor in contrast to the $r_s<r_{1/2}$ region.

With mock data sets, \citet{Bonnivard2015MNRAS.446.3002B} found that the J-Factor was systematically over-estimated without the $r_s > r_{1/2}$ prior (see their Section 4.1). 
Our results for prior dominated systems agree with this result.
This is consistent with galaxy formation simulations, in which galaxies form with $r_s > r_{1/2}$. 
In particular for ultra-faint galaxies, the small $r_s$ values are significantly disfavored by $\Lambda$CDM N-Body simulations. 
For example, the median $r_s$ values for a halos with $V_{\rm max}=5-15 \kms$ are $r_s=120-600 \,{\rm pc}$  \citep{Garrison-Kimmel2014MNRAS.444..222G}.
The systematic over-estimation  in the J-Factor, and  disagreement with N-body simulations, justifies the $r_s > r_{1/2}$ prior.

\subsection{J and D-Factor Compilation}

\begin{figure*}
\includegraphics[width=\textwidth]{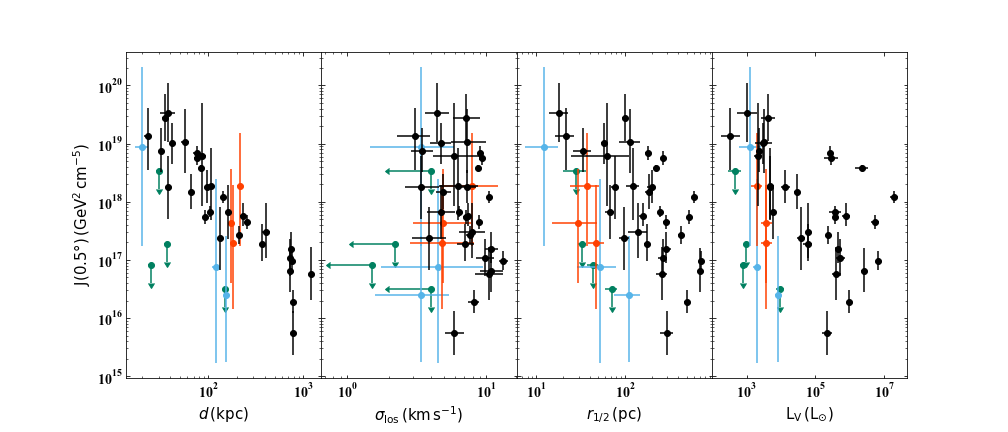}
\caption{J-Factor within a solid angle of 0.5\degree versus (from left to right): distance ($d$; kpc), velocity dispersion ($\slos$; $\kms$), azimuthally averaged half-light radius ($r_{1/2}$; pc), and luminosity (${\rm L_V}$; ${\rm L_{\odot}}$). 
The dSphs are separated based on whether $\slos$ is resolved (see Section~\ref{section:sigma}).  
The groups are resolved (black), unresolved/upper limit (green), large zero-$\slos$ tail (orange), and small  zero-$\slos$ tail (light blue).
Only the peak of the posterior is displayed for the points with zero-$\slos$ tails.
}
\label{fig:jfactor}
\end{figure*}

\begin{figure*}
\includegraphics[width=\textwidth]{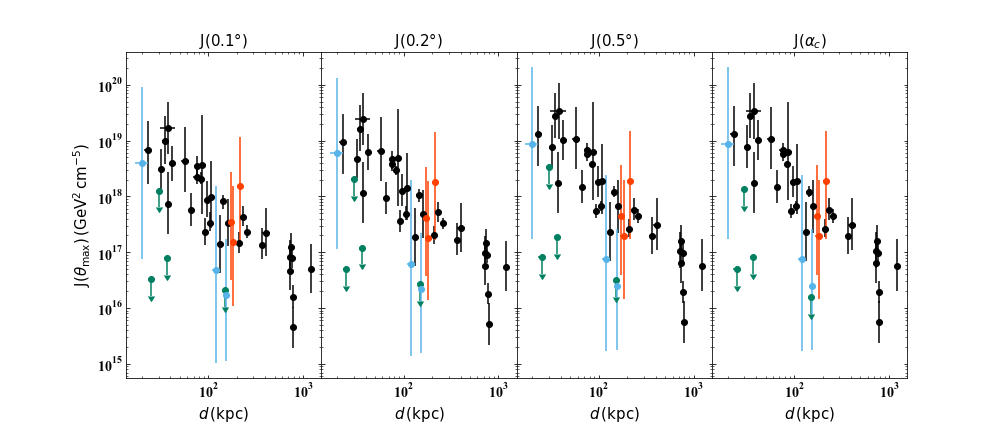}
\caption{J-Factor versus distance for integration angles of 0.1\degree, 0.2\degree, 0.5\degree, and $\alpha_c$. $\alpha_c$ is the angle where the J-Factor uncertainties are minimized \citep{Walker2011ApJ...733L..46W}.}
\label{fig:jfactor_angles}
\end{figure*}

\begin{figure*}
\includegraphics[width=\textwidth]{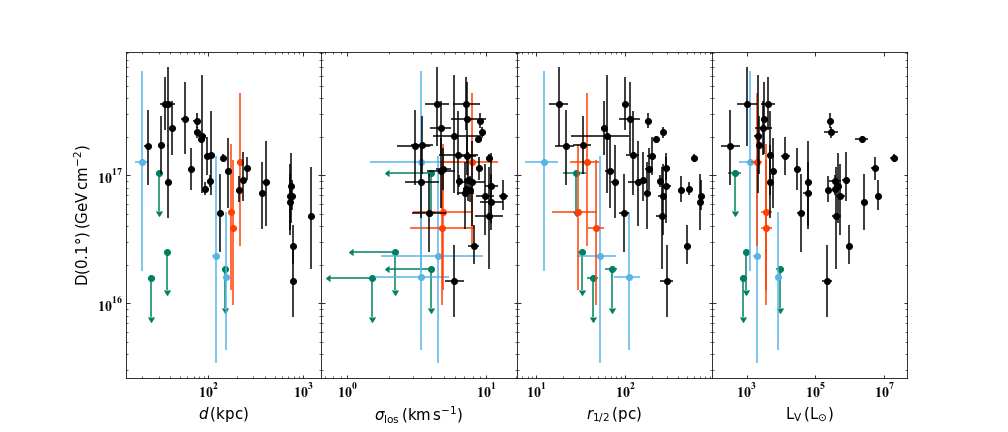}
\caption{Same as Figure~\ref{fig:jfactor} except with the J-Factor replaced with the D-Factor within 0.1\degree.  }
\label{fig:dfactor}
\end{figure*}

We compute the J and D-Factors within solid angles of $\theta_{\rm max}=0.1\degree, 0.2\degree, 0.5\degree,\alpha_c$, where $\alpha_c$ is the angle within which the uncertainties are minimized for J and D-Factor analyses \citep{Walker2011ApJ...733L..46W}.  For the J-Factor, $\alpha_c=2 r_{1/2}/d$, whereas the D-Factor value is half the J-Factor angle, $\alpha_c^D=\alpha_c/2$. 
We provide the first J and D-Factor analysis for the recently discovered MW dSphs Aquarius~II and Pegasus~III. 
We also provide the first J and D-factor analyses for the M31 satellites and a couple of the dispersion supported local group objects (And~XVIII and Cetus). Surprisingly, even though it is at a relatively large distance, And~VII has a J-Factor comparable to some ``faint'' MW satellites and therefore will be useful in future combined gamma-ray likelihood analyses\footnote{In principle most M31 satellites and local group dwarf irregular galaxies could be added to stacked analysis.  This has been discussed within the dynamical framework of rotation curve modeling for local field dwarf irregular galaxies \citep{Gammaldi2017arXiv170601843G}. }.
The J and D-Factor results are compiled in Table~\ref{table:jfac_table}.

Figure~\ref{fig:jfactor}   shows the  summary of the integrated J-Factors within $0.5\degree$~with our sample. From these results we can test how the J-Factor scales with the following observed dSph properties: $d$, $\slos$, $r_{1/2}$, and Visual-band luminosity, ${\rm L_V}$. We separate the dSphs into four categories defined in Section~\ref{section:sigma} based the resolution of $\slos$. 
The intermediate sub-sets have tails in the $\slos$ and J-Factor posterior to small values (equivalent to zero) and we have removed this part of the posterior to compare the ``resolved'' portion to other dSphs (see Section~\ref{section:sigma} for additional details).
Figure~\ref{fig:jfactor_angles} compares the J-Factor at the four different integration angles (0.1\degree, 0.2\degree, 0.5\degree and $\alpha_c$) versus $d$. 

For several of the recently-discovered dSphs in DES (Horologium~I, Reticulum~II, and Tucana~II) the stellar parameters differ between the two independent discovery analyses \citep{Bechtol2015ApJ...807...50B, Koposov2015ApJ...805..130K}.  We compute the J-Factors for these objects with both sets of  stellar parameters (results for both are listed in Table~\ref{data_table}).  
The differences in J-Factor are $\Delta\log_{10}(J(0.5\degree))\sim0.4, 0.1, 0.2$ for the three galaxies respectively.  The largest difference is found in Horologium I where there is a factor of two difference in $r_{1/2}$ between the photometric studies.
This suggests that there is some J-Factor scaling with $r_{1/2}$ despite the lack of an apparently obvious trend in Figure~\ref{fig:jfactor}. 
Throughout the reminder of this analysis we will use J and D-Factor values derived with photometric properties from \citet{Bechtol2015ApJ...807...50B}.
Deeper photometric data will be particularly important to precisely measure the structural parameters for these galaxies.  

In Figure~\ref{fig:dfactor} we show the D-Factor within $0.1\degree$ versus $d$, $\sigma_{\rm los}$, $r_{1/2}$, and ${\rm L_V}$.  
There is an inverse squared scaling with $d$ but it is less pronounced than the trend with J-Factor.  
The more distant systems contain a larger fraction of the dark matter halo within the fixed integration angles and the D-Factor falls off less rapidly than the J-Factor with $\theta_{\rm max}$.

\subsection{J-Factor Scaling}
\label{section:scaling}

\begin{figure*}
\includegraphics[width=\textwidth]{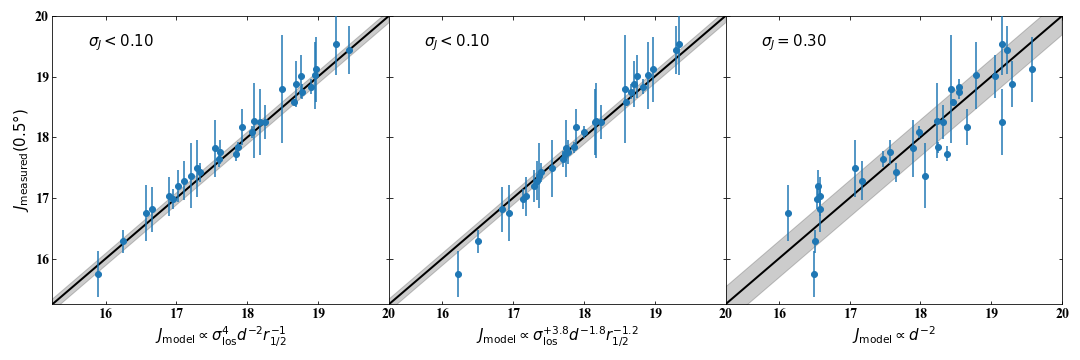}
\caption{J-Factor models versus measured J-Factors at $\theta_{\rm max}=0.5\degree$.  The models from left to right are: $\slos^4 d^{-2} r_{1/2}^{-1}$ (units-based), $\slos^{+3.8} d^{-1.8} r_{1/2}^{-1.2}$ (best-fit), and  $d^{-2}$ (flux-based). For reference we show the one-to-one line with shaded bands set by intrinsic spread of the models  ($\sigma_J$). We list $\sigma_J$ values in the upper-left hand corner.  The units-based and best-fit $\sigma_J$ are the $2-\sigma$ upper limits (95.5\% confidence interval).
}
\label{fig:model}
\end{figure*}

\begin{table*}
\begin{center}
\caption{
Posteriors of J and D-Factor scaling relations.  We list median and 16/84\% confidence intervals.  For upper-limits we quote the $2-\sigma$ value (95.5\% confidence interval). Rows with integer values for a $\gamma$ parameter denote a model fixed to that value. GS15 refers to our results with the \citep{Geringer-Sameth2015ApJ...801...74G} compilation and H16 refers to our results with the \citet{Hutten2016JCAP...09..047H} compilation (see Section~\ref{section:compilation_comparison}).
}
\label{table:scaling}
\begin{tabular}{l cc cc c}
\hline
J(angle)  & $\log_{10}{J_0}$ & $\sigma_J$ & $\gamma_{\slos}$ & $\gamma_d$ & $\gamma_{r_{1/2}}$ \\
\hline
J(0.5\degree) & $17.87_{-0.03}^{+0.04}$ & $<0.10$ & 4 & -2 & -1 \\
J($\alpha_c$) & $17.78\pm0.03$ & $<0.09$ & 4 & -2 & -1 \\
J(0.5\degree) & $18.30\pm0.07$ & $0.30_{-0.06}^{+0.07}$ & 0 & -2 & 0 \\
J($\alpha_c$) & $18.16\pm0.07$ & $0.29_{-0.06}^{+0.07}$ & 0 & -2 & 0 \\
J(0.1\degree) & $17.69_{-0.08}^{+0.09}$ &  $<0.13$ &  $3.8_{-0.5}^{+0.6}$ & $-1.6\pm0.1$ &  $ -1.2\pm0.2$ \\
J(0.2\degree) & $17.84\pm0.09$ &  $<0.12$ &  $3.8\pm0.5$ & $-1.7\pm0.1$ &  $ -1.2\pm0.2$ \\
J(0.5\degree) & $17.96\pm0.9$ &  $<0.10$ &  $3.8\pm0.4$ & $-1.8\pm0.1$ &  $ -1.2\pm0.2$ \\
J($\alpha_c$) & $17.74\pm0.08$ & $<0.10$ & $3.8\pm0.4$ & $-2.0\pm0.1$ & $-0.8\pm0.2$ \\
\hline
\multicolumn{6}{c}{Comparison to other J-Factor Compilations, Section~\ref{section:compilation_comparison}}\\
\hline
J(0.5\degree) GS15  & $18.09\pm0.12$ &  $<0.18$ & $+3.8_{-0.7}^{+0.6}$ & $-1.7\pm0.2$ & $-1.4\pm0.2$ \\
J(0.5\degree) H16 & $18.40\pm0.26$ &  $0.23_{-0.10}^{+0.12}$ & $+2.5\pm1.1$ & $-1.9_{-0.4}^{+0.5}$ & $-1.4\pm0.4$ \\
\hline
\multicolumn{6}{c}{D-Factor Scaling, Section~\ref{section:d_factor}}\\
\hline
D(angle)  & $\log_{10}{D_0}$ & $\sigma_D$ & $\gamma_{\slos}$ & $\gamma_d$ & $\gamma_{r_{1/2}}$ \\
\hline
D($\alpha_c/2$) & $16.57\pm0.02$ & $<0.06$ & 2 & -2 & +1 \\
D(0.1\degree) & $16.98\pm0.05$ &  $<0.06$ &  $1.9\pm0.2$ & $-0.5\pm0.1$ &  $ -0.5\pm0.1$ \\
D(0.2\degree) & $17.41\pm0.06$ &  $<0.08$ &  $1.8\pm0.3$ & $-0.7\pm0.1$ &  $ -0.5\pm0.1$ \\
D(0.5\degree) & $17.93_{-0.10}^{+0.09}$ &  $<0.13$ &  $1.7\pm0.5$ & $-0.9\pm0.2$ &  $ -0.5\pm0.2$ \\
D($\alpha_c/2$) & $16.65\pm0.04$ &  $<0.05$ &  $1.9\pm0.2$ & $-1.9\pm0.6$ &  $ +0.9\pm0.1$ \\
\hline
\multicolumn{6}{c}{Luminosity J-Factor Scaling, Section~\ref{section:luminosity}}\\
\hline
J(angle)  & $\log_{10}{J_0}$ & $\sigma_J$ & $\gamma_{L_V}$ & $\gamma_d$ & $\gamma_{r_{1/2}}$ \\
\hline
J(0.5\degree)  & $18.17\pm0.11$ &  $0.27_{-0.06}^{+0.07}$ & $0.23_{-0.12}^{+0.11}$ & -2 & $-0.5\pm0.4$ \\
J($\alpha_c$)  & $17.96\pm0.11$ &  $0.26_{-0.06}^{+0.07}$ & $0.22_{-0.11}^{+0.10}$ & -2 & $-0.3_{-0.3}^{+0.4}$ \\
\hline
\end{tabular}
\end{center}
\end{table*}
\label{section:luminosity}

\begin{figure*}
\includegraphics[width=\textwidth]{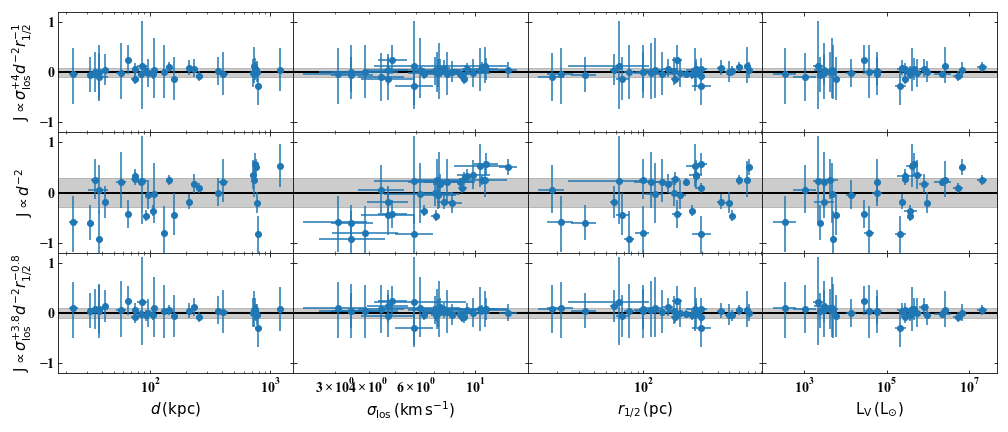}
\caption{The residuals ($J_{\rm measured}- J_{\rm model}$; y-axis) with several different J-Factor (with $\theta_{\rm max}=\alpha_c$) models versus distance ($d$), azimuthally average half-light radius ($r_{1/2}$),  velocity dispersion ($\slos$), and luminosity (${\rm L_V}$).  The models are listed on the y-axis and from top to bottom they are: $J\propto \sigma^4 d^{-2} r_{1/2}^{-1}$ (units-based),
$1/d^2$ (flux-based), and $\sigma^{+3.8} d^{-2} r_{1/2}^{-0.8} $ (best fit).
The bands shows intrinsic scatter ($\sigma_J)$ of each model.  For the top and bottom panels the $\sigma_J$ is the $2-\sigma$ upper limit.
The y-axis displays the same total range for each relationship.  
}
\label{fig:scaling}
\end{figure*}

The first searches for  dark matter annihilation into gamma-rays from new dwarf galaxy candidates discovered in DES \citep{Bechtol2015ApJ...807...50B, Koposov2015ApJ...805..130K, Drlica-Wagner2015ApJ...813..109D} used an empirical scaling relationship between the J-Factor and distance to estimate the J-Factor for the new discoveries \citep{Drlica-Wagner2015ApJ...809L...4D, Albert2017ApJ...834..110A}. 
Since the J-Factor is essentially a ``flux,'' it therefore scales as the inverse of distance squared, as observed in the $d$ subplot of Figure~\ref{fig:jfactor}.
The $d$ relation  is written as 
$\log_{10}{\left (J_{\rm pred}(0.5\degree)/ J_0 \right)} = -2 \log_{10}{\left(d / 100 \kpc \right)}$. 
The normalization, $J_0$,  varies based on the  J-Factor compilation and it  ranges between $\log_{10}{J_0} = 18.1-18.4 \,{\rm GeV^{2} \, cm^{-5}}$  \citep{Geringer-Sameth2015ApJ...801...74G, Bonnivard2015MNRAS.453..849B, Martinez2015MNRAS.451.2524M}. 
One of the recently discovered dSphs, Carina~II, contained a significantly lower J-Factor than the distance scaling prediction \citep{Li2018ApJ...857..145L} and led us to explore more general scaling relations.

Guided by the analytic work of \citet{Evans2016PhRvD..93j3512E}, we examined scaling relations of the form: 
\begin{equation}
\label{equation:scaling_model}
J_{\rm model}(\theta_{\rm max})  = J_0 \left(\frac{\slos}{5 \kms}\right)^{\gamma_{\slos}} \left(\frac{d}{100 {\rm kpc}}\right)^{ \gamma_d} \left(\frac{r_{1/2}}{100 {\rm pc}}\right)^{\gamma_{r_{1/2}}}\,.
\end{equation}
\noindent These quantities are all observed and not dependent on the halo model.  
The distance scaling is required as the J-Factor is a flux, the dispersion probes the mass of the galaxy and the half-light radius sets the inferred mass density for a given dispersion.  

We use a likelihood method to determine best fit parameters and derive errors.
The likelihood we used is:
\begin{equation}
-2\ln{\mathcal{L}}= \sum_{i=1}^{N} \frac{\left( J_{\rm model}(\sigma_{{\rm los},\,i}, d_i, r_{1/2,\,i })-J_{i}\right)^2}{\sigma_J^2 + \epsilon_i^2}+\ln{\left[2 \pi (\sigma_J^2 + \epsilon_i^2)\right]}\,.
\end{equation}
\noindent Here $\sigma_J$ is the intrinsic dispersion or spread of the relation and we assume that the $J_0$ and $\sigma_J$ are in $\log_{10}$ space (for the D-Factor relations we instead refer to these parameters as $D_0$ and $\sigma_D$).
The summation is over the number of galaxies with measured J-Factors and resolved $\slos$.
The posteriors of the model parameters ($J_0$, $\gamma_{\slos}$, $\gamma_{d}$, $\gamma_{r_{1/2}}$, $\sigma_J$) were determined with the {\tt emcee}\footnote{\url{http://dfm.io/emcee/current/}} python Markov chain Monte Carlo python package \citep{ForemanMackey2013PASP..125..306F}.
We assumed uniform priors in the following ranges: $12<J_0<23$, $0<\sigma_J<1$, $-5<\gamma_{\slos} <10$, $-5<\gamma_d <5$, and $-5<\gamma_{r_{1/2}} <5$.  In general, the best fit relation will minimize $\sigma_J$. 

We first examine a relation set by a units based argument.
The units of $[J/G^2]$ are $\textrm{[velocity]}^4$/ $\textrm{[length]}^3$; as a $d^{-2}$ is required for a flux the expected units based relation is $J\propto \slos^4/d^2 r_{1/2}$ .
After fixing the slope parameters, $(\gamma_{\slos}, \gamma_d, \gamma_{r_{1/2}})=(4, -2, -1)$, we found small intrinsic scatter ($\sigma_J<0.10$ at 95.5\% confidence interval) for all  angles.

We explore the full parameter range as a cross check.
In Table~\ref{table:scaling}, we list the posterior values for the 4 different J-Factor angles. 
For all 4 angles we find that the power-law slopes, $\gamma$, agree with the units based parameters values within errors.
We find the minimum $\sigma_J$ occurs at $\alpha_c$ as it is dependent on the $r_{1/2}$, the radius where the mass is best estimated for dispersion supported systems \citep{Walker2009ApJ...704.1274W, Wolf2010MNRAS.406.1220W}.  
The errors in the J-Factors are minimized 
at this angle \citep{Walker2011ApJ...733L..46W}.
We note that there are correlations observed in the posteriors of $J_0-\gamma_{\slos}$ and $\gamma_{\slos}-\gamma_{r_{1/2}}$.

In Figure~\ref{fig:model}, we compare the model predicted J-Factors  to the measured J-Factors.   
The unit-based and best-fit model give equivalent results and are much tighter relations compared to the flux-based (distance) model.
In Figure~\ref{fig:scaling} we examine the residuals of these three models compared to the input parameters ($\slos$, $d$, $r_{1/2}$) and the visual luminosity (${\rm L_V}$).   The unit-based and best-fit models do not show any trends versus any of the parameters whereas in the distance model there is a trend with $\slos$ showing the necessity of a $\slos$ scaling.

The best fit power of $\slos$ is large compared to the other parameters
and there is no apparent trend with respect to $\slos$ in Figure~\ref{fig:jfactor}. 
This is due to the small dynamic range of $\slos$ in the dSphs. 
In particular, $\slos$ only varies by an order of magnitude ($3 \lesssim \slos \lesssim 11 \kms$), whereas the other parameters vary by several orders of magnitude  ($20\lesssim d \lesssim 1200 \kpc$ and $20\lesssim r_{1/2}\lesssim 600 \, {\rm pc}$). 
As $r_{1/2}$ has the weakest scaling in the $J$-factor relation, the trend is only marginally observable in Figure~\ref{fig:jfactor} whereas the $d$ scaling is apparent. 
Since the range in $\slos$ between the dSphs is relatively small, the large $\slos$ power is not seen until the other trends are removed. 
We note that the exponent in the $\slos$ scaling has larger uncertainty than the other parameters due to the small dynamic range of $\slos$ values.

We explore subsets of our sample at two angles (0.5\degree, $\alpha_c$) to check the robustness of our results.  The subsets are: MW satellites, systems with a well measured $\slos$ (error of $\slos<1.5\kms$), luminosity based subsets (removing faint galaxies at $\log_{10}{L} =3.5, 4.0, 4.5$, or removing the `brightest' galaxies $\log_{10}{L} <4.5$), the classical systems (pre-SDSS), systems with a minimum spectroscopic sample size (N > 20), galaxies in the \citet{Geringer-Sameth2015ApJ...801...74G} or \citet{Hutten2016JCAP...09..047H} J-Factor compilations, and distant (non-MW) galaxies.
For $J(\theta_{\rm max}=\alpha_c)$ models, the median power-law parameters varied between: $3.54<\gamma_{\slos}<3.90$, $-2.08<\gamma_d<-2.30$, and $-0.60<\gamma_{r_{1/2}}<-0.72$ except for the classical ($\gamma_{\slos}=4.6$) and distant subsets ($\gamma_d=-2.7$).  These ranges of median values cover the errors in the full sample and show that only in extreme subsets are the slopes different.
We find similar results at $J(\theta_{\rm max}=0.d5\degree)$.  The ranges of median values are: $3.34<\gamma_{\slos}<3.84$, $-1.93<\gamma_d<-2.01$, and $-0.80<\gamma_{r_{1/2}}<-1.0$ with similar excepts for the classical and distant subsets.
Increasing the number of dSphs with kinematics and improving the current precision will help determine the exact scaling.

We advocate to use the J-Factor scaling relation with parameters set by the unit-based values.  
The scaling relation at $\theta_{\rm max}=0.5\degree$ with typical dSph parameters is:

\begin{equation}
\frac{J(0.5\degree)}{  {\rm GeV^{2} \, cm^{-5}}} \approx 10^{17.87} \left(\frac{\slos}{5 \kms}\right)^4 \left(\frac{d}{100\, {\rm kpc} } \right)^{-2} \left( \frac{r_{1/2}}{100 \, {\rm pc} } \right)^{-1} \\.
\label{equation:relation}
\end{equation} 

\noindent The intrinsic spread is constrained to small values with these parameters ($\sigma_J<0.10$).  We show in Section~\ref{section:analytic} that this relation can be derived analytically.

\subsection{D-Factor Scaling}
\label{section:d_factor}

We turn now to deriving a D-Factor scaling relation.
Oddly we find that  the D-Factor relations at fixed angles (i.e. 0.1\degree, 0.2\degree, and 0.5\degree) have similar $\gamma$, they differ from the  D($\alpha_c$) results.  Moreover, the fixed angle D-Factor relations do not have the expected $\gamma_d=-2$ `flux' scaling whereas the D($\alpha_c$) scaling relation does.  After fixing $\gamma_d=-2$, the fixed angle D-Factor relations continue to disagree with the D($\alpha_c$) relation and differ from the non fixed version ($\gamma_{\slos}\approx+3.4$, $\gamma_{r_{1/2}}\approx-0.05$). As the the best fit free distance model had a different slope it is no surprise to see the other parameters change. 

The D($\alpha_c/2$) scaling relation is consistent with a unit-based argument; $[D/G]=\textrm{[velocity]}^2/ \textrm{[length]} = \slos^2 r_{1/2}/d^2$.  
Fixing the $\gamma$ parameters to the unit based argument for D($\alpha_c$) results in a slightly larger scatter, $\sigma_D<0.06$, versus $\sigma_D<0.05$ for the free parameters.
The best fit scaling relation with typical dSph parameters with fixed $\gamma$ parameters is: 

\begin{equation}
\frac{D(\alpha_c/2)}{{\rm GeV \, cm^{-2}}} \approx 10^{16.57} \left(\frac{\slos}{5 \kms}\right)^2 \left(\frac{d}{100 {\rm kpc} } \right)^{-2} \left( \frac{r_{1/2}}{100 {\rm pc} }\right)^{+1} \\.
\end{equation} 

\begin{figure}
\includegraphics[width=\columnwidth]{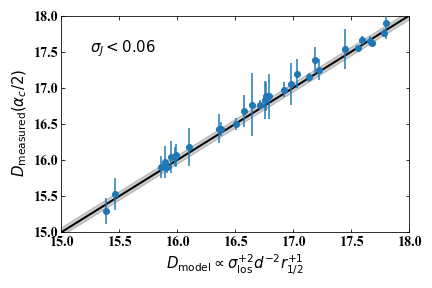}
\caption{Comparison of model predications to measured the D-Factor at $\theta_{\rm max}=\alpha_c/2$.}
\label{fig:dfactor_model}
\end{figure}

\begin{figure*}
\includegraphics[width=\textwidth]{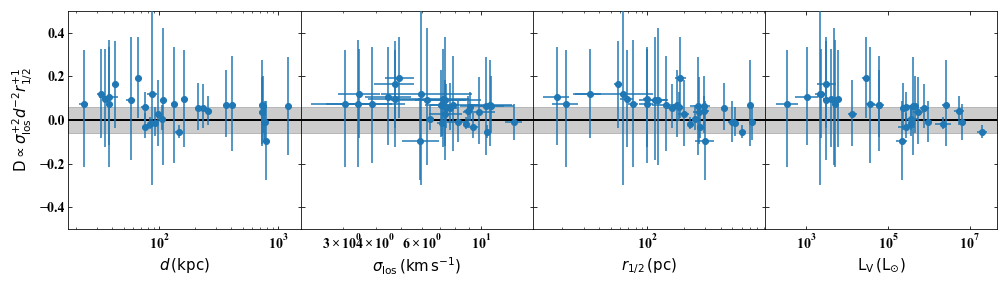}
\caption{Similar to Figure~\ref{fig:scaling}, except with the D-Factor model at $\theta=\alpha_c/2$ instead of J-Factor models.}
\label{fig:dfactor_residual}
\end{figure*}

The fixed angle scaling relation with $\gamma$ set by to the units-based parameters have large scatter ($\sigma_D\approx0.37-0.53$).  The only angle for which a D-Factor scaling relation applies is at $\alpha_c/2$.
In Figures~\ref{fig:dfactor_model},~\ref{fig:dfactor_residual}, we compare the D-Factor measurements to the D-Factor model at $\alpha_c$.  
Figures~\ref{fig:dfactor_model} provides a one-to-one comparison and Figure~\ref{fig:dfactor_residual} compares the residuals to the input parameters ($\slos$, $d$, $r_{1/2}$) and the luminosity as a cross check.  Encouragingly, there are no remaining trends with the input parameters.

The lack of a consistent scaling relationship at fixed versus $\alpha_c/2$ could be due to how the D-Factor scales with $\theta$.
The shape of the D-Factor integrand with respect to $\theta$ (only line-of-sight integration) is quite different compared to the analogous J-Factor integrand.
For the J-Factor integrand the majority of ``signal'' comes within $r_s$ and always decreases with respect to $\theta$.  
For the D-Factor however, the integrand initially increases with respect to $\theta$ and then turns over at $\theta \approx 0.4 r_s/d$.  
There is significantly slower falloff with respect to $\theta$ for the D-Factor than for the J-Factor.  
At a fixed integration angle, the shape of the D-Factor integrand will vary between objects, whereas the variable angle, $\alpha_c/2$, is more likely to probe similar parts of the D-Factor integrand.

\subsection{Analytic Relation}
\label{section:analytic}

We can derive the form of our scaling relation by appealing to the analytic work of \citet{Evans2016PhRvD..93j3512E}. They derive  analytic J and D-Factors for several simple halo profiles including the NFW profile. Their analytic J and D-Factors contain two, generally valid, simplifying assumptions: first, the dwarf is distant enough to simplify the angular part of the integration (projection from infinite distance versus finite distance) and second, the  dark matter halo has no truncation radius (infinite $r_t$). 
We find that their formula works remarkably well for the NFW J-Factor; generally, the percent error between the numerical integration and approximate analytic calculation in our posterior distribution is $\leq0.1\%$.  
The D-Factor formula however, preforms quite poorly due to the infinite tidal radius assumption (percent errors range range from $1-50\%$).  As the D-Factor has a larger dependence on the total size of the dark matter halo than the J-Factor, the approximation tends to over estimate the D-Factor.  We therefore only focus on the analytic J-Factor work.

The bulk of our derivation is in Appendix~\ref{appendix:jfactor}.
Briefly, we start with the analytic J-Factor formula for the NFW profile in \citet[Equation 16]{Evans2016PhRvD..93j3512E} and replace the halo scale density  with observed quantities ($\slos$, $r_{1/2}$) using the half-mass estimators \citep{Walker2009ApJ...704.1274W,Wolf2010MNRAS.406.1220W}.
At $\alpha_c$, the J-Factor can then be written as:
\begin{equation}
J(\alpha_c) =  \frac{\slos^4}{G^2 d^2 r_{1/2}} F(r_{1/2}/r_s),
\label{equation:analytic_sec3}
\end{equation}
\noindent where the $F$ is an analytic function derived in the appendix.
The $\slos^4$ dependence comes from the half-mass estimators ($J\propto \rho_s^2 \propto M^2 \propto \slos^4$) and $d^{-2}$ dependence from the ``flux'' nature of the J-Factor.
The remaining unit is 1/[length] and implies that $J\propto 1/r_{1/2}$.
The first part of Equation~\ref{equation:analytic_sec3} is the scaling relation we find while the remainder is dependent on the ratio $r_{1/2}/r_s$.  
With $\langle r_{1/2}/r_s \rangle \approx 0.25$, Equation~\ref{equation:analytic_sec3} has the same normalization as our scaling relations for $\alpha_c$ (see Table~\ref{table:scaling}).  

\subsection{Comparison with other J-Factor Compilations}
\label{section:compilation_comparison}

As a cross check we apply our methodology to the compilations\footnote{In both cases we exclude galaxies (Draco~II, Leo~IV, Leo~V, Pisces~II, Segue~2, and Triangulum~II) that were excluded from our sample.}  of \citet{Geringer-Sameth2015ApJ...801...74G} and \citet{Hutten2016JCAP...09..047H}.  
Both works  tabulated  the $d$ and $r_{1/2}$ values used in their analysis, however, neither compiled $\slos$ for their samples.  We substitute our own calculations of $\slos$ as the kinematic samples for most galaxies overlap. The \citet{Geringer-Sameth2015ApJ...801...74G} analysis used a more generalized dark matter halo (double power-law) and assume the same stellar anisotropy and stellar density profile as our work.
The \citet{Hutten2016JCAP...09..047H}\footnote{Most of the analysis in this compilation was initially presented in \citet{Bonnivard2015MNRAS.453..849B}} study in contrast used more general profiles for the dark matter, stellar anisotropy, and stellar light compared to our work. 
Both provide J-Factors at 0.5\degree.

We find $J_0=18.09\pm0.12$, $\sigma_J=<0.18$, $\gamma_{\slos}=3.8_{-0.7}^{+0.6}$, $\gamma_d=-1.7\pm0.2$, and $\gamma_{r_{1/2}}=-1.4\pm0.2$ 
for the \citet{Geringer-Sameth2015ApJ...801...74G} compilation and 
find  $J_0=18.4\pm0.26$, $\sigma_J=0.23_{-0.10}^{+0.12}$, $\gamma_{\slos}=2.5_{-1.1}^{+1.1}$, $\gamma_d=-1.9_{-0.4}^{+0.5}$, and $\gamma_{r_{1/2}}=-1.4\pm0.4$ for the \citet{Hutten2016JCAP...09..047H} compilation.
Both studies have a larger normalizations than our study which may be caused by the generalized halo model and they have slightly steeper values for the $\gamma_{r_{1/2}}$ parameter.  
The best fit  slope parameters with the \citet{Geringer-Sameth2015ApJ...801...74G} compilation agree with the analytic and units based argument. 
In contrast, we find with the \citet{Hutten2016JCAP...09..047H} compilation a smaller $\gamma_{\slos}$ value with much larger errors and a large intrinsic scatter ($\sigma_J$).  
As a cross check we examined our sample with the subset of galaxies in each of these compilations and found our results were consistent.
Overall, the results with the \citet{Geringer-Sameth2015ApJ...801...74G} compilation are consistent with our compilation while the \citet{Hutten2016JCAP...09..047H} compilation results are consistent within $1-\sigma$.

\subsection{Scaling with Luminosity}
\label{section:luminosity}

In the era of deep and wide surveys, it will become increasingly more difficult for the measurement of stellar kinematics  within individual systems to keep up with the rate at which new systems are discovered.
So it is important to determine how the J-factor scales with parameters other than $\slos$. For satellites without stellar kinematics, the stellar luminosity, ${\rm L_V}$, may be a potential replacement for $\slos$ in the J-factor scaling relation. In our dSph sample, there is a rough correlation between $\slos$ and ${\rm L_V}$.
We explore J-factor scaling relations replacing ${\rm L_V}$ for $\slos$.
The best-fit relation  we find is (fixing $\gamma_d=-2$):
\begin{equation}
\frac{J(0.5\degree )}{{\rm GeV^{2} \, cm^{-5}}} \approx 10^{18.17} \left(\frac{\rm L_V}{10^4 L_{\odot}}\right)^{0.23} \left(\frac{d}{100\, {\rm kpc} } \right)^{-2} \left( \frac{r_{1/2}}{100 \, {\rm pc} } \right)^{-0.5}  \\.
\label{eq:relationLv}
\end{equation}
\noindent This scaling relation has an intrinsic scatter of $\sigma_J=0.27$, which is larger than the scaling with $\slos$ ($\sigma_J<0.10$), but it is equivalent in scatter to a simple  $d^{-2}$ scaling ($\sigma_J=0.30$).
Similar results are found for $J(\alpha_c)$. 
We note there is a clear anti-correlation between the parameters, $\gamma_{\rm L_V}$-$\gamma_{r_{1/2}}$.  We provide the best fit parameters for these two angles in Table~\ref{table:scaling}.

\begin{figure}
\includegraphics[width=\columnwidth]{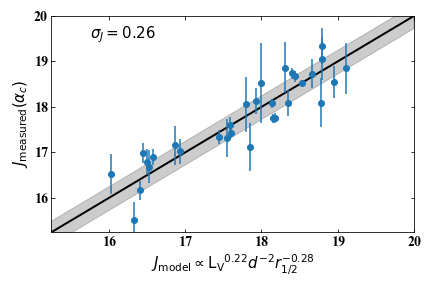}
\caption{Best-fit J-Factor luminosity model compared to the measurements at $\theta_{\rm max}=\alpha_c$. The shaded band represents the intrinsic scatter of the model ($\sigma_J$) and is noted in the upper-left. }
\label{fig:jfac_lum_model}
\end{figure}

\begin{figure*}
\includegraphics[width=\textwidth]{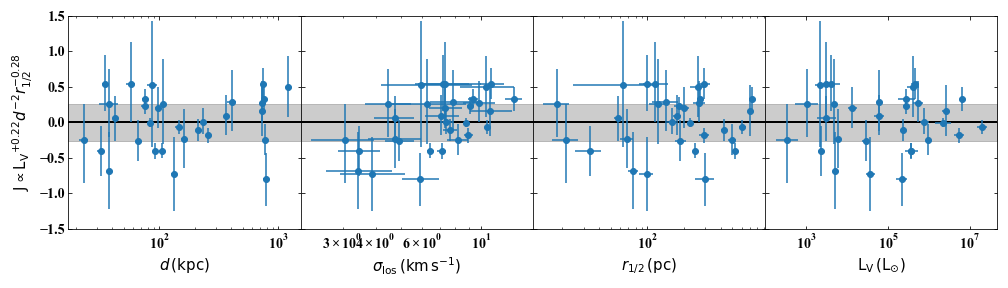}
\caption{Similar to Figure~\ref{fig:scaling}, except with best-fit J-Factor luminosity models at $\theta_{\rm max}=\alpha_c$. There is still a non-zero slope in the $\slos$ residual. }
\label{fig:lum_jfactor_residual}
\end{figure*}

In Figures~\ref{fig:jfac_lum_model},~\ref{fig:lum_jfactor_residual} we provide comparison figures for the luminosity models. 
In Figure~\ref{fig:lum_jfactor_residual} compares the model residuals versus the model inputs ($d$, $r_{1/2}$, ${\rm L_V}$) and $\slos$.  
In the residuals there is a trend versus $\slos$ implying that the best fit model should include $\slos$.
As the luminosity scaling has larger scatter compared to our $\slos$ scaling relation this is not surprising.

We explored different subsets of our sample (similar to Section~\ref{section:scaling}) to check the robustness of the luminosity based scaling relation.  
Most subsets contain similar results with  similar $\sigma_J$.
The subsets with large differences   are the classical, distant, and `faint' ($\log_{10}{L_V}<4.5$) dSph  subsets.  All of these subsets have small sample sizes and  show the same anti-correlation between $\gamma_{\rm L_V}$-$\gamma_{r_{1/2}}$.  However, the `faint' galaxies subset significantly differs ($\gamma_{\rm L_V}\approx-0.46$,  $\gamma_{r_{1/2}}\approx+0.5$).

There are a couple caveats to note with the above results. Many of the M31 dSphs with smaller kinematic samples do not follow the same ${\rm L_V}-\slos$ trend as the  dSphs in our sample \citep{Tollerud2012ApJ...752...45T, Collins2013ApJ...768..172C}. The recently discovered diffuse galaxy, Crater II, also falls below this relationship~\citep{Torrealba2016MNRAS.459.2370T, Caldwell2017ApJ...839...20C}. However, our $\slos$ relationship predicts $\log_{10}{J(0.5\degree)}= 15.6$, while the published measurement (at a larger angle) is  $\log_{10}{J(1.4\degree)}= 15.7$.

Given the present data, we make predictions with the  $J-{\rm L_V}$ relation for galaxy candidates without kinematics in Table~\ref{table:predictions}. Comparable predictions with the distance scaling relation can be found in Table 1 of \citet{Albert2017ApJ...834..110A}.  
The scatter we find in the ${\rm L_V}-r_{1/2}-d$ relation is marginally better than $d^{-2}$.  It is possible that additional and more precise J-Factors of faint systems can improve predictions for J-Factors for systems without kinematic measurements.

\begin{table}
\caption{
Predictions for galaxies without kinematics based on Equation~\ref{eq:relationLv}.
Citations:
(a) \citep{Drlica-Wagner2015ApJ...813..109D}
(b) \citep{Homma2018PASJ...70S..18H}
(c) \citep{Carlin2017AJ....154..267C}
(d) \citep{Kim2015ApJ...808L..39K}
(e) \citep{Drlica-Wagner2016ApJ...833L...5D}
(f) \citep{Bechtol2015ApJ...807...50B}
(g) \citep{Laevens2015ApJ...813...44L}
}
\label{table:predictions}
\begin{tabular}{l cc cc c}
\hline
Galaxy & ${\rm L_V}$ & $r_{1/2}$ & $d$ & J(0.5\degree) & Citation\\
& $L_{\odot}$ & pc & kpc & $\junit$ \\
\hline
Cetus II  & 8.6e1 & 17 & 30 & 19.1 & a \\
Cetus III  & 8.2e2 & 44 & 251 & 17.3 & b\\
Columba I  & 4.1e3 & 98 & 183 & 17.6 & c\\
Grus II  & 3.1e3 & 93 & 53 & 18.4 & a\\ 
Horologium II  & 94e2 &  33 & 78 & 18.4 & d\\
Indus II  & 4.5e3 & 181 & 214 & 17.3 & a\\ 
Pictor I  & 2.6e3 & 43 & 126 & 18.0 & f\\
Pictor II  & 1.6e3 & 46 & 45 & 18.9 & e\\
Phoenix II  & 26e3 & 33 & 95 & 18.3 & f\\
Reticulum III  & 1.8e3 & 64 & 92 & 18.2 & a\\
Sagittarius II  & 1.0e4 & 33 & 67 & 18.8 & g\\
Tucana IV  & 2.1e3 & 98 &  48 & 18.7 & a\\
Tucana V  & 3.7e2 &  9.3 &  55 & 18.9 & a\\
Virgo I  & 1.2e2 & 30 & 91 & 18.1 & b\\
\hline
\end{tabular}
\end{table}

\subsection{Limitations}

\par As emphasized above, to obtain the J and D-factors, we have implemented the standard spherical Jeans-based likelihood analysis. We have chosen this method for the simplicity in interpretation of the results, and to perform a fair comparison with previous authors who have performed similar analyses. 

We note, however, that there are some limitations to our likelihood model.  First we have fixed to the dark matter halo profile to be an NFW profile. Several works have presented generalized dark matter profiles \citep[e.g.][]{Geringer-Sameth2015ApJ...801...74G, Bonnivard2015MNRAS.453..849B}. When allowing for a more flexible model for the dark matter halo, the general trend is toward increasing the J-factor, though there are no substantial biases in the results. 
A second limitation may be in our parameterization of the stellar velocity anisotropy profile. The APOSTLE hydrodynamical simulations find stellar anisotropy profiles that can be well approximated as constant profiles \citep{Campbell2017MNRAS.469.2335C}.
~\citet{Chiappo2017MNRAS.466..669C} have studied in particular the different between constant and Osipkov-Merrit anisotropy profiles in J-Factor analysis.
Given the small data sets for many of the dSphs we have studied, our assumptions for stellar anisotropy and stellar density profiles are adequate. This is particularly true for ultra-faints, though for the brightest dSphs a more flexible model may be ultimately required.

We have also assumed spherical symmetry for our dynamical models. Deviations from spherical symmetry have been studied in previous J-factors analysis, including the use of axisymmetric Jeans modeling~\citep{Hayashi2016MNRAS.461.2914H} and made-to-measure models  \citep{Sanders2016PhRvD..94f3521S}. It is also worth noting that incorrect dSph membership can affect the inferred velocity dispersion and therefore J-Factor in ultra-faints \citep{Bonnivard2016MNRAS.462..223B, Ichikawa2017arXiv170605481I}, though this is less of a problem in the bright classical satellites.  We have modified some commonly used spectroscopic data sets, mostly removing RR Lyrae stars (dSph members but the stars are variable in velocity).
Binary stars can inflate the velocity dispersion of dSphs and therefore inflate the J-Factors \citep{Minor2010ApJ...721.1142M, McConnachie2010ApJ...722L.209M, Kirby2017ApJ...838...83K, Spencer2017AJ....153..254S}.  If binary stars have inflated the velocity dispersion, then our scaling relations imply that the J and D-Factors will be simply scaled down $\left(\sigma_{\rm true}/\sigma_{\rm inflated}\right)^\alpha$ with $\alpha=4,2$ respectively.  
Inflated $\slos$ values could potentially affect the derived normalizations ($J_0/D_0$) for our scaling relations but this requires future multi-epoch data to test.

\section{Conclusion}
\label{section:conclusion}

\par In this paper we have presented a compilation of J and D-Factors for MW, M31, and LF dSphs. In addition to adding new J and D-factor calculations for newly-discovered MW satellites (Aquarius II and Pegasus III), we provide the first calculations for satellites of M31 and within the local field between the MW and M31. From this compilation, we derive scaling relations for the J and D-Factors. We find a scaling relation for the J-Factor to be: $J \propto \slos^4/d^2/r_{1/2}$ and find the respective relation for the D-Factor to be: $D \propto \slos^2 r_{1/2}/d^2$. The strongest scaling for the J-Factor is based on $\slos$ or dynamical mass but due to the small relative range of $\slos$ for the MW satellites ($\slos \approx 2-13\kms$) this scaling is hard to pick out over the distance scaling ($d\approx 20-450 {\rm kpc}$). We further find that there is no strong scaling with the stellar luminosity. 

\par While performing a dynamical modeling as outlined in Section~\ref{section:method} is ideal for computing the J and D-Factors, for small data samples and for systems with unresolved velocity dispersions this full analysis may not be feasible. It is in these instances in which the scaling relations are particularly important to provide an estimate of the J and D-factors.

\par Improved galaxy kinematics in the ultra-low luminosity region is required to ultimately test the scaling relations we have discussed. However, with current observational facilities, it is likely difficult to significantly improve upon the stellar kinematics of known dSphs, especially the faintest-known systems. It would be particularly interesting if a larger data sample reveals a scaling between the luminosity and the J or D-factors. Though the scaling with $L_V$ does not do better than pure distance scaling given the current data, if more data were to reveal a $J-L_V$ scaling, such a relation would be especially useful to estimate the J-Factor for galaxies before they are followed up spectroscopically. This will likely be especially true in the LSST era, during which many low luminosity dSphs are predicted to be discovered \citep[e.g.][]{Hargis2014ApJ...795L..13H} and spectroscopic characterization and dark matter properties may be difficult before 30-meter class facilities become available.  Future multi-object spectrographs on 10-meter class telescopes such as the Maunakea Spectroscopic Explorer \citep{McConnachie2016arXiv160600060M, McConnachie2016arXiv160600043M} and the Prime Focus Spectrograph on Subaru \citep{Takada2014PASJ...66R...1T} will significantly improve follow-up kinematic surveys and therefore J-Factors of current and future dwarf galaxies.  For future spectroscopic follow-up of dwarfs, our scaling relation are useful for selecting the objects  that  will maximize improvements in the indirect detection of dark matter.

\section*{Acknowledgments}

We thank Josh Simon and Matt Walker for providing spectroscopic data. LES acknowledges support from DOE Grant de-sc0010813. We acknowledge generous support from the Texas A\&M University and the George P. and Cynthia Woods Institute for Fundamental Physics and Astronomy.  We thank the referee for their careful reading of the paper that improved the presentation and strengthened the conclusions of the paper.

Databases and software: 
Python packages: \texttt{Astropy}\footnote{\url{http://www.astropy.org}} \citep{astropy2013},  \texttt{NumPy} \citep{numpy}, \texttt{iPython} \citep{ipython}, \texttt{SciPy} \citep{scipy}, \texttt{matplotlib} \citep{matplotlib}, {\tt emcee}\citep{ForemanMackey2013PASP..125..306F}, and \texttt{corner}\citep{corner}.  This research has made use of NASA's Astrophysics Data System Bibliographic Services.

Posterior chain and additional plots can be found at the following webpage \url{https://github.com/apace7/J-Factor-Scaling}.

\bibliographystyle{mnras}

\bibliography{linked_bib_file}

\appendix

\section{Analytic Derivation of the J-Factor Scaling Relation}
\label{appendix:jfactor}
We begin with the analytic form of the NFW J-Factor  \citep[Equation 16 of][]{Evans2016PhRvD..93j3512E}:

\begin{align}
J(\theta) &= \frac{\pi \rho_s^2 r_s^3}{3 d^2 } \frac{1}{\Delta^2} \left[2 y \left( 7y - 4y^3 + 3 \pi \Delta^4 \right) \right. \nonumber \\
&+ \left. 6 \left(2 \Delta^6 -2 \Delta^2 - y^4\right) \frac{1}{\Delta} {\rm Arcsech}(y)\right] \nonumber \\
J(\theta) &= \frac{\pi \rho_s^2 r_s^3}{3 d^2 } f(d \theta /r_s) \label{equation:evans}
\end{align}

\noindent where $y = d \theta/r_s$ and  $\Delta^2 = 1 - y^2$.  For brevity we will write the functional portion of Equation~\ref{equation:evans} as $f(d\theta/r_s)$. 
For the NFW mass profile:

\begin{equation}
\label{equation:nfw_mhalf}
\rho_s r_s = \frac{M_{1/2}}{4 \pi r_s^2 g(r_{1/2}/r_s)} = \frac{5 \slos^2 r_{1/2}}{2 G}\frac{1}{4 \pi r_s^2 g(r_{1/2}/r_s)}\, \\
\end{equation}

\noindent where $g(x) = \log{\left( 1+x \right)} - x/(1+x)$ and the later equality follows from the half-mass estimators \citep{Walker2009ApJ...704.1274W,Wolf2010MNRAS.406.1220W}.

Replacing $\rho_s$ in Equation~\ref{equation:evans} with Equation~\ref{equation:nfw_mhalf} and examining $\theta=\alpha_c = 2 r_{1/2}/d$ we find:

\begin{align*}
J(\alpha_c) & = \frac{\pi r_s}{3 d^2} \left(\frac{5 \slos^2 r_{1/2}}{8 \pi {\rm G} r_s^2} \frac{1}{g(r_{1/2}/r_s)} \right)^2 f(2 r_{1/2}/r_s)\\
& = \frac{25}{192 \pi} 
\frac{\slos^4}{{\rm G^2} r_{1/2} d^2 }
\left( \frac{r_{1/2}}{r_s}\right)^3 
\frac{f(2 r_{1/2}/r_s)}{\left[g(r_{1/2}/r_s \right]^2} \\
& = \frac{\slos^4}{{\rm G^2} r_{1/2} d^2 } F(r_{1/2}/r_s)  \\ 
\end{align*}

\noindent In the last line we have moved all the numerical constants and the $r_{1/2}/r_s$ dependence into a single function.
If we assume $r_{1/2}/r_s\approx 0.25$,  we find the same normalization as our scaling relation.

\onecolumn

\begin{landscape}
\begin{longtable}{lc cc cc c cc cccc}
\caption{
Errors are $\pm 1 \sigma$ for original study.
Photometric properties, velocity dispersion, and spectroscopic source for all objects in our study.  Description of columns: (1) Name of galaxy. (2) Heliocentric distance ($d$; kpc). (3) Azimuthally averaged half-light radius ($r_{1/2}$; pc). (4) Velocity dispersion ($\slos$; $\kms$).  (5) Number of spectroscopic members including data source. (6) V-Band absolute magnitude. 
The distance column error bars reflect numbers quoted in the original study and  $r_{1/2}$ error bars include errors on distance, major-axis half-light radius, and ellipticity.
Upper limits are quoted at 99.5\% confidence interval.
Letters in columns refer to the following citations:
(a) \citep{Kuehn2008ApJ...674L..81K}
(b) \citep{Martin2008ApJ...684.1075M}
(c) \citep{Simon2007ApJ...670..313S}
(d) \citep{Karczmarek2015AJ....150...90K}
(e) \citep{Munoz2018ApJ...860...66M}
(f) \citep{Walker2009AJ....137.3100W}
(g) \citep{Bonanos2004AJ....127..861B}
(h) \citep{Walker2015MNRAS.448.2717W}
(i) \citep{Pietrzynski2009AJ....138..459P}
(j) \citep{Bate2015MNRAS.453..690B}
(k) \citep{Stetson2014PASP..126..616S}
(l) \citep{Smolcic2007AJ....134.1901S}
(m) \citep{Mateo2008ApJ...675..201M}
(n) \citep{Bellazzini2005MNRAS.360..185B}
(o) \citep{Coleman2007AJ....134.1938C}
(p) \citep{Spencer2017ApJ...836..202S}
(q) \citep{Martinez-Vazquez2015MNRAS.454.1509M}
(r) \citep{Okamoto2017MNRAS.467..208O}
(s) \citep{Cicuendez2017arXiv170904519C}
(t) \citep{Bellazzini2002AJ....124.3222B}
(u) M. Spencer et al., in prep 
(v) \citep{Torrealba2016MNRAS.463..712T}
(w) \citep{DallOra2006ApJ...653L.109D}
(x) \citep{Koposov2011ApJ...736..146K}
(y) \citep{Greco2008ApJ...675L..73G}
(z) \citep{Sand2012ApJ...756...79S}
(aa) \citep{Torrealba2018MNRAS.tmp..170T}
(ab) \citep{Li2018ApJ...857..145L}
(ac) \citep{Musella2009ApJ...695L..83M}
(ad) \citep{Munoz2010AJ....140..138M}
(ae) \citep{Laevens2015ApJ...813...44L}
(af) \citep{Martin2016MNRAS.458L..59M}
(ag) \citep{Koposov2015ApJ...805..130K}
(ah) \citep{Walker2016ApJ...819...53W}
(ai) \citep{Musella2012ApJ...756..121M}
(aj) \citep{Coleman2007ApJ...668L..43C}
(ak) \citep{Bechtol2015ApJ...807...50B}
(al) \citep{Koposov2015ApJ...811...62K}
(am) \citep{Vivas2016AJ....151..118V}
(an) \citep{Martin2015ApJ...804L...5M}
(ao) \citep{Kirby2015ApJ...810...56K}
(ap) \citep{Moretti2009ApJ...699L.125M}
(aq) \citep{Okamoto2012ApJ...744...96O}
(ar) \citep{Medina2017ApJ...845L..10M}
(as) \citep{Collins2017MNRAS.467..573C}
(at) \citep{Kim2016ApJ...833...16K}
(au) \citep{Simon2015ApJ...808...95S}
(av) \citep{Belokurov2007ApJ...654..897B}
(aw) \citep{Simon2011ApJ...733...46S}
(ax) \citep{Boettcher2013AJ....146...94B}
(ay) \citep{Belokurov2009MNRAS.397.1748B}
(az) \citep{Kirby2013ApJ...770...16K}
(ba) \citep{Laevens2015ApJ...802L..18L}
(bb) \citep{Kirby2017ApJ...838...83K}
(bc) \citep{Drlica-Wagner2015ApJ...813..109D}
(bd) \citep{Simon2017ApJ...838...11S}
(be) \citep{Garofalo2013ApJ...767...62G}
(bf) \citep{Okamoto2008A_A...487..103O}
(bg) \citep{DallOra2012ApJ...752...42D}
(bh) \citep{Willman2005AJ....129.2692W}
(bi) \citep{Willman2011AJ....142..128W}
(bj) \citep{Bernard2009ApJ...699.1742B}
(bk) \citep{McConnachie2006MNRAS.365.1263M}
(bl) \citep{Kirby2014MNRAS.439.1015K}
(bm) \citep{Crnojevic2016ApJ...824L..14C}
(bn) \citep{Li2017ApJ...838....8L}
(bo) \citep{deJong2008ApJ...680.1112D}
(bp) \citep{Conn2012ApJ...758...11C}
(bq) \citep{Martin2016ApJ...833..167M}
(br) \citep{Tollerud2012ApJ...752...45T}
(bs) \citep{McConnachie2005MNRAS.356..979M}
}\\
\label{data_table}
\endfirsthead
\endhead
\hline
Galaxy & Distance & $r_{1/2}$ & $\sigma$ & N & $M_V$    \\
& kpc & pc & $\kms$ & & & \\
\hline
Canes Venatici I & 210.0 $\pm$ 6.0(a) & $423\pm25$(b) & $7.6_{-0.4}^{+0.5}$ & 209(c) & -8.6 $\pm$ 0.15\\
Carina & 105.6 $\pm$ 5.4(d) & $248\pm13$(e) & $6.4_{-0.2}^{+0.2}$ & 758(f) & -9.1 $\pm$ 0.4\\
Draco & 76.0 $\pm$ 6.0(g) & $182\pm15$(b) & $9.1_{-0.3}^{+0.3}$ & 476(h) & -8.75 $\pm$ 0.15\\
Fornax & 147.0 $\pm$ 9.0(i) & $602\pm37$(j) & $10.6_{-0.2}^{+0.2}$ & 2409(f) & -13.4 $\pm$ 0.3\\
Leo I & 258.2 $\pm$ 9.5(k) & $292\pm26$(l) & $9.0_{-0.4}^{+0.4}$ & 327(m) & -12.0 $\pm$ 0.3\\
Leo II & 233.0 $\pm$ 15.0(n) & $158\pm14$(o) & $7.4_{-0.4}^{+0.4}$ & 175(p) & -9.9 $\pm$ 0.3\\
Sculptor & 83.9 $\pm$ 1.5(q) & $223\pm5$(e) & $8.8_{-0.2}^{+0.2}$ & 1349(f) & -11.04 $\pm$ 0.5\\
Sextans & 92.5 $\pm$ 2.2(r) & $523\pm23$(s) & $7.1_{-0.3}^{+0.3}$ & 424(f) & -9.1 $\pm$ 0.1\\
Ursa Minor & 76.0 $\pm$ 4.0(t) & $269\pm15$(e) & $9.3_{-0.4}^{+0.4}$ & 311(u) & -8.8 $\pm$ 0.5\\
\hline
Aquarius II & 107.9 $\pm$ 3.3(v) & $123\pm21$(v) & $6.2_{-1.7}^{+2.6}$ & 9(v) & -4.36 $\pm$ 0.14\\
Bo\"{o}tes I & 66.0 $\pm$ 3.0(w) & $187\pm19$(b) & $4.9_{-0.6}^{+0.7}$ & 37(x) & -6.3 $\pm$ 0.2\\
Canes Venatici II & 160.0 $\pm$ 7.0(y) & $68\pm8$(z) & $4.7_{-1.0}^{+1.2}$ & 25(c) & -4.6 $\pm$ 0.2\\
Carina II & 37.4 $\pm$ 0.4(aa) & $77\pm8$(aa) & $3.4_{-0.8}^{+1.2}$ & 14(ab) & -4.4 $\pm$ 0.1\\
Coma Berenices & 42.0 $\pm$ 1.5(ac) & $57\pm4$(ad) & $4.7_{-0.8}^{+0.9}$ & 58(c) & -3.9 $\pm$ 0.6\\
Draco II & 20.0 $\pm$ 3.0(ae) & $12\pm5$(ae) & $3.4_{-1.9}^{+2.5}$ & 9(af) & -2.9 $\pm$ 0.8\\
Grus I & 120.2 $\pm$ 11.1(ag) & $52\pm26$(ag) & $4.5_{-2.8}^{+5.0}$ & 5(ah) & -3.4 $\pm$ 0.3\\
Hercules & 132.0 $\pm$ 6.0(ai) & $98\pm13$(aj) & $3.9_{-1.0}^{+1.3}$ & 30(c) & -6.6 $\pm$ 0.3\\
Horologium I B & 87.0 $\pm$ 8.0(ak) & $63\pm44$(ak) & $5.9_{-1.8}^{+3.3}$ & 5(al) & -3.5 $\pm$ 0.3\\
Horologium I K & 79.0 $\pm$ 7.0(ag) & $32\pm5$(ag) & $5.9_{-1.8}^{+3.3}$ & 5(al) & -3.4 $\pm$ 0.1\\
Hydra II & 151.0 $\pm$ 8.0(am) & $71\pm11$(an) & $<6.82$ & 13(ao) & -5.1 $\pm$ 0.3\\
Leo IV & 154.0 $\pm$ 5.0(ap) & $111\pm36$(aq) & $3.4_{-1.8}^{+2.0}$ & 17(c) & -4.92 $\pm$ 0.2\\
Leo V & 173.0 $\pm$ 5.0(ar) & $30\pm17$(z) & $4.9_{-1.9}^{+3.0}$ & 8(as) & -4.1 $\pm$ 0.4\\
Pegasus III & 215.0 $\pm$ 12.0(at) & $37\pm14$(at) & $7.9_{-3.1}^{+4.4}$ & 7(at) & -3.4 $\pm$ 0.4\\
Pisces II & 183.0 $\pm$ 15.0(z) & $48\pm10$(z) & $4.8_{-2.0}^{+3.3}$ & 7(ao) & -4.1 $\pm$ 0.4\\
Reticulum II B & 32.0 $\pm$ 2.0(ak) & $34\pm8$(ak) & $3.4_{-0.6}^{+0.7}$ & 25(au) & -3.6 $\pm$ 0.1\\
Reticulum II K & 30.0 $\pm$ 2.0(ag) & $32\pm3$(ag) & $3.4_{-0.6}^{+0.7}$ & 25(au) & -2.7 $\pm$ 0.1\\
Segue 1 & 23.0 $\pm$ 2.0(av) & $21\pm5$(b) & $3.1_{-0.8}^{+0.9}$ & 62(aw) & -1.5 $\pm$ 0.7\\
Segue 2 & 36.6 $\pm$ 2.45(ax) & $33\pm3$(ay) & $<3.20$ & 25(az) & -2.6 $\pm$ 0.1\\
Triangulum II & 30.0 $\pm$ 2.0(ba) & $28\pm8$(ba) & $<6.36$ & 13(bb) & -1.8 $\pm$ 0.5\\
Tucana II K & 57.5 $\pm$ 5.3(ag) & $162\pm35$(ag) & $7.3_{-1.7}^{+2.6}$ & 10(ah) & -3.8 $\pm$ 0.1\\
Tucana II B & 57.5 $\pm$ 5.3(ak) & $115\pm32$(ak) & $7.3_{-1.7}^{+2.6}$ & 10(ah) & -3.9 $\pm$ 0.2\\
Tucana III & 25.0 $\pm$ 2.0(bc) & $43\pm6$(bc) & $<2.18$ & 26(bd) & -2.4 $\pm$ 0.2\\
Ursa Major I & 97.3 $\pm$ 5.85(be) & $199\pm21$(bf) & $7.3_{-1.0}^{+1.2}$ & 36(c) & -5.5 $\pm$ 0.3\\
Ursa Major II & 34.7 $\pm$ 2.1(bg) & $99\pm7$(ad) & $7.2_{-1.4}^{+1.8}$ & 19(c) & -4.2 $\pm$ 0.5\\
Willman 1 & 38.0 $\pm$ 7.0(bh) & $18\pm4$(b) & $4.5_{-0.8}^{+1.0}$ & 40(bi) & -2.7 $\pm$ 0.7\\
\hline
Cetus & 780.0 $\pm$ 40.0(bj) & $498\pm37$(bk) & $8.2_{-0.7}^{+0.8}$ & 116(bl) & -10.1 $\pm$ 0.0\\
Eridanus II & 366.0 $\pm$ 17.0(bm) & $176\pm14$(bm) & $7.1_{-0.9}^{+1.2}$ & 28(bn) & -7.1 $\pm$ 0.3\\
Leo T & 407.0 $\pm$ 38.0(bo) & $142\pm35$(b) & $7.9_{-1.5}^{+2.0}$ & 19(c) & -7.1 $\pm$ 0.0\\
And I & 727.0 $\pm$ 17.5(bp) & $699\pm28$(bq) & $10.9_{-1.7}^{+2.3}$ & 51(br) & -11.2 $\pm$ 0.2\\
And III & 723.0 $\pm$ 21.0(bp) & $265\pm31$(bq) & $9.8_{-1.3}^{+1.5}$ & 62(br) & -9.5 $\pm$ 0.3\\
And V & 742.0 $\pm$ 21.5(bp) & $294\pm33$(bq) & $11.0_{-1.0}^{+1.2}$ & 85(br) & -9.3 $\pm$ 0.2\\
And VII & 763.0 $\pm$ 35.0(bs) & $717\pm40$(bk) & $13.3_{-1.0}^{+1.0}$ & 136(br) & -12.2 $\pm$ 0.0\\
And XIV & 793.0 $\pm$ 50.0(bp) & $296\pm52$(bq) & $5.9_{-0.9}^{+1.0}$ & 48(br) & -8.5 $\pm$ 0.35\\
And XVIII & 1214.0 $\pm$ 41.5(bp) & $260\pm37$(bq) & $10.5_{-2.1}^{+2.8}$ & 22(br) & -9.2 $\pm$ 0.35\\
\hline
\end{longtable}
\end{landscape}

\twocolumn

\onecolumn

\begin{landscape}
\begin{longtable}{lc cc cc c cc cccc}
\caption{
J and D-factor results for all objects in our study.  Description of columns: (1) Name of galaxy. (2) $\alpha_c\approx2 r_{1/2}/d$ (degree) \citep{Walker2011ApJ...733L..46W} (3-6) J-Factor ($\log_{10}$; $\junit$)within angular cone of  $\theta_{\rm max}=0.1\degree,0.2\degree  0.5\degree, \alpha_c$. (7-10) D-Factor ($\log_{10}$; $\dunit$)within angular cone of  $\theta_{\rm max}=0.1\degree, 0.2\degree, 0.5\degree, \alpha_c/2$.
We list the median value and error bars refer to the $15.87\%$ and $84.13\%$ confidence intervals.  Upper limits are quoted at 99.5\% confidence interval. 
For Horologium I, Reticulum II, and Tucana II, the first and second results refer to parameters with \citet{Bechtol2015ApJ...807...50B} and \citet{Koposov2015ApJ...805..130K} respectively.
}\\
\label{table:jfac_table}
\endfirsthead
\endhead
\hline
Galaxy & $\alpha_c$ & ${\rm J} (0.1\degree)$ &   ${\rm J} (0.2\degree)$ & ${\rm J} (0.5\degree)$ & ${\rm J} (\alpha_c)$ &${\rm D} (0.1\degree)$ &   ${\rm D} (0.2\degree)$ &   ${\rm D} (0.5\degree)$ & ${\rm D} (\alpha_c/2)$  \\
&  deg & $\junit$ & $\junit$& $\junit$& $\junit$& $\dunit$& $\dunit$& $\dunit$& $\dunit$\\
\hline
Canes Venatici I & 0.231 & $17.16_{-0.18}^{+0.19}$ & $17.31_{-0.15}^{+0.16}$ & $17.42_{-0.15}^{+0.17}$ & $17.33_{-0.15}^{+0.15}$ & $16.89_{-0.09}^{+0.11}$ & $17.31_{-0.15}^{+0.16}$ & $17.79_{-0.27}^{+0.26}$ & $16.98_{-0.10}^{+0.12}$\\
Carina & 0.269 & $17.53_{-0.15}^{+0.18}$ & $17.69_{-0.11}^{+0.13}$ & $17.83_{-0.09}^{+0.10}$ & $17.75_{-0.10}^{+0.11}$ & $16.95_{-0.04}^{+0.05}$ & $17.39_{-0.07}^{+0.09}$ & $17.88_{-0.15}^{+0.18}$ & $17.14_{-0.05}^{+0.06}$\\
Draco & 0.276 & $18.35_{-0.13}^{+0.15}$ & $18.58_{-0.12}^{+0.13}$ & $18.83_{-0.12}^{+0.12}$ & $18.68_{-0.12}^{+0.12}$ & $17.42_{-0.06}^{+0.06}$ & $17.92_{-0.08}^{+0.08}$ & $18.54_{-0.14}^{+0.11}$ & $17.66_{-0.07}^{+0.07}$\\
Fornax & 0.476 & $17.91_{-0.14}^{+0.12}$ & $18.02_{-0.12}^{+0.11}$ & $18.09_{-0.10}^{+0.10}$ & $18.08_{-0.10}^{+0.10}$ & $17.14_{-0.03}^{+0.03}$ & $17.54_{-0.03}^{+0.03}$ & $17.97_{-0.04}^{+0.05}$ & $17.63_{-0.03}^{+0.03}$\\
Leo I & 0.130 & $17.36_{-0.11}^{+0.12}$ & $17.52_{-0.10}^{+0.10}$ & $17.64_{-0.12}^{+0.14}$ & $17.43_{-0.10}^{+0.11}$ & $17.06_{-0.09}^{+0.09}$ & $17.50_{-0.16}^{+0.13}$ & $18.01_{-0.28}^{+0.20}$ & $16.76_{-0.07}^{+0.07}$\\
Leo II & 0.078 & $17.64_{-0.16}^{+0.18}$ & $17.71_{-0.17}^{+0.18}$ & $17.76_{-0.18}^{+0.22}$ & $17.60_{-0.17}^{+0.19}$ & $16.97_{-0.17}^{+0.21}$ & $17.30_{-0.24}^{+0.32}$ & $17.64_{-0.33}^{+0.50}$ & $16.43_{-0.09}^{+0.11}$\\
Sculptor & 0.305 & $18.32_{-0.11}^{+0.11}$ & $18.47_{-0.08}^{+0.08}$ & $18.58_{-0.05}^{+0.05}$ & $18.53_{-0.06}^{+0.07}$ & $17.28_{-0.02}^{+0.02}$ & $17.71_{-0.03}^{+0.04}$ & $18.20_{-0.08}^{+0.09}$ & $17.55_{-0.03}^{+0.03}$\\
Sextans & 0.650 & $17.37_{-0.24}^{+0.24}$ & $17.56_{-0.19}^{+0.19}$ & $17.73_{-0.12}^{+0.13}$ & $17.77_{-0.11}^{+0.11}$ & $16.90_{-0.05}^{+0.05}$ & $17.36_{-0.04}^{+0.05}$ & $17.90_{-0.09}^{+0.11}$ & $17.65_{-0.06}^{+0.07}$\\
Ursa Minor & 0.409 & $18.54_{-0.23}^{+0.18}$ & $18.66_{-0.17}^{+0.15}$ & $18.75_{-0.12}^{+0.12}$ & $18.74_{-0.12}^{+0.13}$ & $17.34_{-0.04}^{+0.04}$ & $17.75_{-0.05}^{+0.06}$ & $18.20_{-0.08}^{+0.14}$ & $17.76_{-0.05}^{+0.06}$\\
\hline
Aquarius II & 0.131 & $18.00_{-0.57}^{+0.64}$ & $18.14_{-0.57}^{+0.63}$ & $18.27_{-0.58}^{+0.66}$ & $18.06_{-0.57}^{+0.63}$ & $17.16_{-0.31}^{+0.35}$ & $17.58_{-0.37}^{+0.38}$ & $18.07_{-0.50}^{+0.47}$ & $16.88_{-0.30}^{+0.33}$\\
Bo\"{o}tes I & 0.325 & $17.76_{-0.29}^{+0.31}$ & $17.96_{-0.28}^{+0.30}$ & $18.17_{-0.29}^{+0.31}$ & $18.08_{-0.28}^{+0.30}$ & $17.05_{-0.16}^{+0.17}$ & $17.52_{-0.19}^{+0.20}$ & $18.11_{-0.30}^{+0.25}$ & $17.38_{-0.18}^{+0.19}$\\
Canes Venatici II & 0.049 & $17.52_{-0.41}^{+0.43}$ & $17.68_{-0.43}^{+0.44}$ & $17.82_{-0.47}^{+0.47}$ & $17.31_{-0.41}^{+0.43}$ & $17.04_{-0.29}^{+0.26}$ & $17.48_{-0.37}^{+0.29}$ & $18.01_{-0.51}^{+0.36}$ & $16.05_{-0.22}^{+0.22}$\\
Carina II & 0.234 & $17.87_{-0.55}^{+0.56}$ & $18.06_{-0.53}^{+0.55}$ & $18.25_{-0.54}^{+0.55}$ & $18.09_{-0.53}^{+0.55}$ & $16.95_{-0.28}^{+0.29}$ & $17.41_{-0.31}^{+0.32}$ & $17.95_{-0.40}^{+0.38}$ & $17.06_{-0.29}^{+0.29}$\\
Coma Berenices & 0.157 & $18.59_{-0.32}^{+0.32}$ & $18.79_{-0.32}^{+0.32}$ & $19.00_{-0.35}^{+0.36}$ & $18.73_{-0.32}^{+0.32}$ & $17.37_{-0.21}^{+0.21}$ & $17.85_{-0.29}^{+0.24}$ & $18.45_{-0.44}^{+0.29}$ & $17.20_{-0.20}^{+0.20}$\\
Draco II$^{*}$ & 0.071 & $18.60_{-1.72}^{+1.36}$ & $18.77_{-1.71}^{+1.34}$ & $18.93_{-1.70}^{+1.39}$ & $18.51_{-1.72}^{+1.35}$ & $17.11_{-0.85}^{+0.71}$ & $17.53_{-0.86}^{+0.75}$ & $18.02_{-0.87}^{+0.84}$ & $16.42_{-0.85}^{+0.68}$\\
Grus I$^{*}$ & 0.050 & $16.68_{-1.66}^{+1.50}$ & $16.79_{-1.65}^{+1.51}$ & $16.88_{-1.66}^{+1.51}$ & $16.52_{-1.66}^{+1.49}$ & $16.37_{-0.84}^{+0.79}$ & $16.71_{-0.85}^{+0.83}$ & $17.00_{-0.86}^{+0.87}$ & $15.52_{-0.83}^{+0.75}$\\
Hercules & 0.086 & $17.15_{-0.52}^{+0.52}$ & $17.27_{-0.52}^{+0.51}$ & $17.37_{-0.53}^{+0.53}$ & $17.12_{-0.52}^{+0.53}$ & $16.71_{-0.31}^{+0.30}$ & $17.09_{-0.37}^{+0.37}$ & $17.52_{-0.51}^{+0.50}$ & $16.18_{-0.27}^{+0.26}$\\
Horologium I B & 0.084 & $18.56_{-0.87}^{+0.89}$ & $18.69_{-0.87}^{+0.88}$ & $18.79_{-0.86}^{+0.90}$ & $18.52_{-0.87}^{+0.90}$ & $17.31_{-0.45}^{+0.48}$ & $17.69_{-0.50}^{+0.54}$ & $18.14_{-0.63}^{+0.65}$ & $16.77_{-0.42}^{+0.45}$\\
Horologium I K & 0.046 & $19.01_{-0.64}^{+0.76}$ & $19.14_{-0.66}^{+0.76}$ & $19.27_{-0.71}^{+0.77}$ & $18.82_{-0.63}^{+0.77}$ & $17.55_{-0.47}^{+0.43}$ & $17.95_{-0.58}^{+0.49}$ & $18.40_{-0.78}^{+0.62}$ & $16.58_{-0.33}^{+0.38}$\\
Hydra II & 0.054 & $ < 17.57$ & $ < 17.69$ & $ < 17.77$ & $ < 17.40$ & $ < 16.93$ & $ < 17.36$ & $ < 17.62$ & $ < 16.05$\\
Leo IV$^{*}$ & 0.083 & $16.24_{-1.18}^{+1.00}$ & $16.35_{-1.16}^{+1.00}$ & $16.40_{-1.15}^{+1.01}$ & $16.19_{-1.17}^{+1.02}$ & $16.21_{-0.57}^{+0.51}$ & $16.55_{-0.57}^{+0.52}$ & $16.74_{-0.56}^{+0.55}$ & $15.68_{-0.57}^{+0.50}$\\
Leo V$^{*}$ & 0.020 & $17.54_{-1.02}^{+0.90}$ & $17.62_{-1.02}^{+0.91}$ & $17.65_{-1.03}^{+0.91}$ & $17.20_{-1.01}^{+0.96}$ & $16.72_{-0.60}^{+0.53}$ & $17.00_{-0.67}^{+0.61}$ & $17.13_{-0.72}^{+0.65}$ & $15.35_{-0.50}^{+0.45}$\\
Pegasus III$^{*}$ & 0.020 & $18.21_{-0.95}^{+0.86}$ & $18.29_{-0.97}^{+0.88}$ & $18.30_{-0.97}^{+0.89}$ & $17.89_{-0.98}^{+0.86}$ & $17.13_{-0.62}^{+0.52}$ & $17.41_{-0.69}^{+0.59}$ & $17.47_{-0.72}^{+0.61}$ & $15.74_{-0.48}^{+0.43}$\\
Pisces II$^{*}$ & 0.030 & $17.19_{-1.09}^{+1.00}$ & $17.27_{-1.09}^{+1.00}$ & $17.30_{-1.09}^{+1.00}$ & $16.96_{-1.11}^{+1.02}$ & $16.60_{-0.58}^{+0.53}$ & $16.89_{-0.64}^{+0.57}$ & $17.01_{-0.66}^{+0.61}$ & $15.48_{-0.54}^{+0.50}$\\
Reticulum II B & 0.121 & $18.48_{-0.35}^{+0.36}$ & $18.68_{-0.34}^{+0.36}$ & $18.88_{-0.37}^{+0.39}$ & $18.54_{-0.35}^{+0.36}$ & $17.24_{-0.23}^{+0.23}$ & $17.71_{-0.32}^{+0.27}$ & $18.30_{-0.50}^{+0.33}$ & $16.89_{-0.20}^{+0.21}$\\
Reticulum II K & 0.121 & $18.55_{-0.33}^{+0.35}$ & $18.75_{-0.33}^{+0.35}$ & $18.96_{-0.37}^{+0.38}$ & $18.61_{-0.33}^{+0.35}$ & $17.27_{-0.23}^{+0.22}$ & $17.75_{-0.31}^{+0.25}$ & $18.34_{-0.49}^{+0.32}$ & $16.91_{-0.20}^{+0.20}$\\
Segue 1 & 0.107 & $18.83_{-0.61}^{+0.52}$ & $18.98_{-0.58}^{+0.49}$ & $19.12_{-0.58}^{+0.49}$ & $18.85_{-0.60}^{+0.51}$ & $17.23_{-0.31}^{+0.28}$ & $17.62_{-0.36}^{+0.36}$ & $18.08_{-0.49}^{+0.53}$ & $16.83_{-0.29}^{+0.25}$\\
Segue 2 & 0.103 & $ < 17.80$ & $ < 18.01$ & $ < 18.21$ & $ < 17.81$ & $ < 16.88$ & $ < 17.38$ & $ < 17.98$ & $ < 16.42$\\
Triangulum II & 0.109 & $ < 19.27$ & $ < 19.48$ & $ < 19.73$ & $ < 19.30$ & $ < 17.64$ & $ < 18.12$ & $ < 18.72$ & $ < 17.19$\\
Tucana II K & 0.326 & $18.41_{-0.51}^{+0.58}$ & $18.62_{-0.50}^{+0.55}$ & $18.84_{-0.50}^{+0.55}$ & $18.74_{-0.49}^{+0.55}$ & $17.36_{-0.25}^{+0.27}$ & $17.83_{-0.27}^{+0.29}$ & $18.41_{-0.33}^{+0.32}$ & $17.69_{-0.26}^{+0.28}$\\
Tucana II B & 0.232 & $18.63_{-0.56}^{+0.61}$ & $18.82_{-0.54}^{+0.59}$ & $19.02_{-0.53}^{+0.58}$ & $18.86_{-0.54}^{+0.59}$ & $17.44_{-0.27}^{+0.29}$ & $17.90_{-0.29}^{+0.31}$ & $18.46_{-0.36}^{+0.35}$ & $17.54_{-0.27}^{+0.29}$\\
Tucana III & 0.199 & $ < 17.31$ & $ < 17.50$ & $ < 17.71$ & $ < 17.50$ & $ < 16.58$ & $ < 17.06$ & $ < 17.69$ & $ < 16.58$\\
Ursa Major I & 0.235 & $17.94_{-0.31}^{+0.34}$ & $18.10_{-0.28}^{+0.30}$ & $18.26_{-0.27}^{+0.29}$ & $18.14_{-0.28}^{+0.29}$ & $17.15_{-0.15}^{+0.15}$ & $17.59_{-0.18}^{+0.18}$ & $18.10_{-0.29}^{+0.28}$ & $17.26_{-0.15}^{+0.15}$\\
Ursa Major II & 0.327 & $18.99_{-0.41}^{+0.45}$ & $19.21_{-0.39}^{+0.43}$ & $19.44_{-0.39}^{+0.41}$ & $19.34_{-0.39}^{+0.41}$ & $17.56_{-0.20}^{+0.21}$ & $18.04_{-0.23}^{+0.23}$ & $18.64_{-0.31}^{+0.28}$ & $17.90_{-0.22}^{+0.23}$\\
Willman 1 & 0.056 & $19.22_{-0.46}^{+0.47}$ & $19.38_{-0.46}^{+0.47}$ & $19.53_{-0.50}^{+0.50}$ & $19.06_{-0.47}^{+0.49}$ & $17.56_{-0.32}^{+0.30}$ & $17.99_{-0.45}^{+0.36}$ & $18.52_{-0.68}^{+0.47}$ & $16.68_{-0.22}^{+0.23}$\\
\hline
Cetus & 0.073 & $16.20_{-0.20}^{+0.21}$ & $16.26_{-0.19}^{+0.19}$ & $16.28_{-0.19}^{+0.20}$ & $16.17_{-0.21}^{+0.22}$ & $16.45_{-0.14}^{+0.17}$ & $16.74_{-0.20}^{+0.26}$ & $17.03_{-0.26}^{+0.39}$ & $15.91_{-0.09}^{+0.10}$\\
Eridanus II & 0.055 & $17.13_{-0.26}^{+0.29}$ & $17.22_{-0.28}^{+0.31}$ & $17.28_{-0.31}^{+0.34}$ & $17.03_{-0.26}^{+0.29}$ & $16.86_{-0.28}^{+0.25}$ & $17.22_{-0.40}^{+0.33}$ & $17.60_{-0.54}^{+0.47}$ & $16.06_{-0.15}^{+0.16}$\\
Leo T & 0.040 & $17.34_{-0.41}^{+0.45}$ & $17.43_{-0.43}^{+0.46}$ & $17.49_{-0.45}^{+0.49}$ & $17.16_{-0.41}^{+0.45}$ & $16.95_{-0.34}^{+0.32}$ & $17.30_{-0.46}^{+0.40}$ & $17.67_{-0.60}^{+0.53}$ & $15.97_{-0.21}^{+0.23}$\\
And I & 0.110 & $16.67_{-0.35}^{+0.38}$ & $16.76_{-0.35}^{+0.38}$ & $16.81_{-0.36}^{+0.40}$ & $16.69_{-0.35}^{+0.38}$ & $16.79_{-0.21}^{+0.22}$ & $17.15_{-0.26}^{+0.27}$ & $17.51_{-0.38}^{+0.35}$ & $16.44_{-0.19}^{+0.21}$\\
And III & 0.042 & $16.92_{-0.27}^{+0.29}$ & $16.99_{-0.29}^{+0.31}$ & $17.03_{-0.30}^{+0.33}$ & $16.78_{-0.27}^{+0.30}$ & $16.84_{-0.30}^{+0.26}$ & $17.17_{-0.42}^{+0.33}$ & $17.49_{-0.56}^{+0.47}$ & $15.90_{-0.15}^{+0.16}$\\
And V & 0.045 & $17.10_{-0.21}^{+0.24}$ & $17.16_{-0.23}^{+0.25}$ & $17.19_{-0.25}^{+0.29}$ & $16.99_{-0.21}^{+0.23}$ & $16.92_{-0.29}^{+0.26}$ & $17.23_{-0.39}^{+0.35}$ & $17.54_{-0.51}^{+0.50}$ & $16.05_{-0.13}^{+0.14}$\\
And VII & 0.108 & $16.89_{-0.17}^{+0.17}$ & $16.95_{-0.15}^{+0.16}$ & $16.98_{-0.15}^{+0.16}$ & $16.90_{-0.16}^{+0.17}$ & $16.84_{-0.11}^{+0.13}$ & $17.15_{-0.15}^{+0.22}$ & $17.46_{-0.21}^{+0.34}$ & $16.50_{-0.08}^{+0.09}$\\
And XIV & 0.043 & $15.65_{-0.36}^{+0.36}$ & $15.71_{-0.36}^{+0.37}$ & $15.74_{-0.37}^{+0.38}$ & $15.53_{-0.38}^{+0.37}$ & $16.18_{-0.28}^{+0.28}$ & $16.49_{-0.37}^{+0.36}$ & $16.79_{-0.51}^{+0.50}$ & $15.29_{-0.18}^{+0.18}$\\
And XVIII & 0.025 & $16.70_{-0.43}^{+0.44}$ & $16.74_{-0.44}^{+0.46}$ & $16.75_{-0.44}^{+0.48}$ & $16.53_{-0.42}^{+0.44}$ & $16.68_{-0.42}^{+0.38}$ & $16.95_{-0.50}^{+0.48}$ & $17.21_{-0.59}^{+0.60}$ & $15.53_{-0.22}^{+0.23}$\\
\hline
\end{longtable}
$^{*}$ Galaxy contains tail in posterior ($\slos$ and $J/D$) distributions.  Values should be used with caution. 
\end{landscape}

\twocolumn

\bsp	
\label{lastpage}
\end{document}